\documentclass[useAMS,usenatbib]{mnras}
\usepackage{hyperref}
\usepackage{epsf}
\usepackage{amssymb}
\usepackage{graphicx}
\usepackage{eqnarray}
\usepackage{verbatim}
\usepackage{multirow}
\usepackage{amsmath}
\usepackage{float}
\usepackage[usenames]{color}

\newcommand {\bc}{\begin {center}}
\newcommand {\ec}{\end {center}}

\def\g {{\rm g}}

\def\s {{\rm s}}
\def\i {{\rm i}}

\def\cm {{\rm cm}}
\def\km {{\rm km}}
\def\Mpc {{\rm Mpc}}

\title[X-ray emission from WHIM]{X-ray emission from warm-hot intergalactic medium: \\ the role of resonantly scattered cosmic X-ray background}
\author[Khabibullin \& Churazov]{I.~Khabibullin$^{1,2}$, E.~Churazov$^{1,2}$\\
\centerline{}\\
$^{1}$ MPI f\"ur Astrophysik, Karl-Schwarzschild Str. 1, Garching D-85741, Germany\\
$^{2}$ Space Research Institute, Profsoyuznaya str. 84/32, Moscow, 117997, Russia\\
}
\begin{document}
\label{firstpage}
\date{Accepted 2018 October 30. Received 2018 October 19; in original form 2018 August 19.}
\pagerange{\pageref{firstpage}--\pageref{lastpage}} \pubyear{2018}
\maketitle
\begin{abstract}

We revisit calculations of the X-ray emission from the warm-hot intergalactic medium (WHIM) with particular focus on contribution from the resonantly scattered cosmic X-ray background (CXB). If the significant part of the CXB emission is resolved into point sources, the properties of the WHIM along the line of sight are recorded in the absorption lines in the stacked spectrum of resolved sources \textit{and} in the emission lines in the remaining diffuse signal. For the strongest resonant lines, this implies a factor of $\sim30$ boost in emissivity compared to the intrinsic emissivity over the major part of the density-temperature parameter space region relevant for WHIM. The overall boost for the 0.5-1 keV band is $\sim4$, declining steeply at temperatures above $10^{6}$ K and over-densities $\delta\gtrsim100$. In addition to the emissivity boost, contribution of the resonant scattering changes relative intensities of the lines, so it should be taken into account when line-ratio-diagnostics from high resolution spectra or redshift determination from low resolution spectra are considered. Comparison between WHIM signatures in X-ray absorption and emission should allow differentiating truly diffuse gas of small overdensity from denser clumps having small filling factor by future X-ray missions.
 
\end{abstract}
\section{Introduction}
\label{s:introduction}

~~~~~~ Gravitational collapse of over-dense regions, proceeding via unilateral compression into sheets followed by formation of filaments and compact virialized objects (\citealt{1970A&A.....5...84Z}; see \citealt{2012ARA&A..50..353K} for a recent review), provides steady heating to the baryonic  content of the Universe \citep{1972A&A....20..189S}, so that approximately half of its mass attains temperature in between $10^5$ K and $10^7$ K, and some $\sim 15$ \% gets virialized and heated to $10^7$ K, at $z\lesssim 1$ \citep{1999ApJ...514....1C,2001ApJ...552..473D,2006ApJ...650..560C}. The latter portion constitutes the baryonic content of massive galaxy clusters and groups, i.e. intracluster medium (ICM), and it is readily observed via intense thermal X-ray emission and the Sunyaev-Zel'dovich distortion of the Cosmic Microwave Background (CMB, e.g. \citealt{1981ASPRv...1....1S}), thanks to a combination of high temperature and relatively high density ($\gtrsim 200$ the critical density). 

Detection of the former portion, the warm-hot intergalactic medium (WHIM), however, turns out to be extremely challenging, since its thermal emission is not only weak, due to relatively low density ($\sim 10$ the mean density), but also falls into extreme ultraviolet range, that is both heavily absorbed and contaminated by the interstellar medium (ISM) of our own Galaxy \citep[e.g.][]{2007ARA&A..45..221B}. Besides that, being progressively heated to temperatures comparable to or higher than the hydrogen ionization potential, and also photoionized by the cosmic UV and X-ray background radiation, the WHIM matter at $z\lesssim 1$ lacks sufficient amount of neutral hydrogen to be seen via Ly$\alpha$ absorption in spectra of distant quasars, as the matter at higher redshifts ($z\gtrsim2$) does (\citealt{1998ARA&A..36..267R,2013AJ....145...69L}; see \citealt{2009RvMP...81.1405M} and \citealt{2016ARA&A..54..313M} for recent reviews). As a result, the dominant role of WHIM in the baryonic budget of the Universe at $z\sim0$ is most clearly demonstrated by inability of the all observationally accounted baryons in the Local Universe to amount more than a half of $\Omega_{b}$, measured by the combination of CMB observations with the primordial nucleosynthesis models \citep{1998ApJ...503..518F,2016A&A...594A..13P}.  

Fortunately, by $z\sim0$, WHIM is already substantially enriched by metals expelled from the star-forming galaxies via intensive galactic-scale outflows, so that it is characterized by typical metallicity $Z\sim0.1Z_{\odot}$ \citep[e.g.][]{2001ApJ...560..599A,2009MNRAS.399..574W}. For gas temperature $\lesssim 10^7$ K, some of the most abundant metals like oxygen and iron stay not fully-ionized, even when photo-ionization by cosmic X-ray background radiation (CXB) is taken into account \citep{2016MNRAS.459..310R}. As a result, X-ray emission of WHIM at such temperatures is dominated by line emission from these atoms \citep{2003PASJ...55..879Y,2010MNRAS.407..544B}. Additionally, thanks to resonant scattering in these lines, WHIM should also reveal itself in absorption against bright background UV/X-ray sources {\citep[e.g.][and references therein]{2007ARA&A..45..221B,2008SSRv..134...25R,2018Natur.558..406N}}. The latter possibility is of particular importance for detection of the gas with the lowest density, since the absorption amplitude is proportional to the column density of the intervining matter (as far as it doesn't become saturated), i.e. it scales linearly with the gas number density, while the thermal emission is proportional to the emission measure, so it scales as the gas density squared \citep[e.g.][]{2008SSRv..134..405P}.

\begin{figure}
\centering
\label{fig:sketch}
\includegraphics[bb=50 220 600 680,width=1.0\columnwidth]{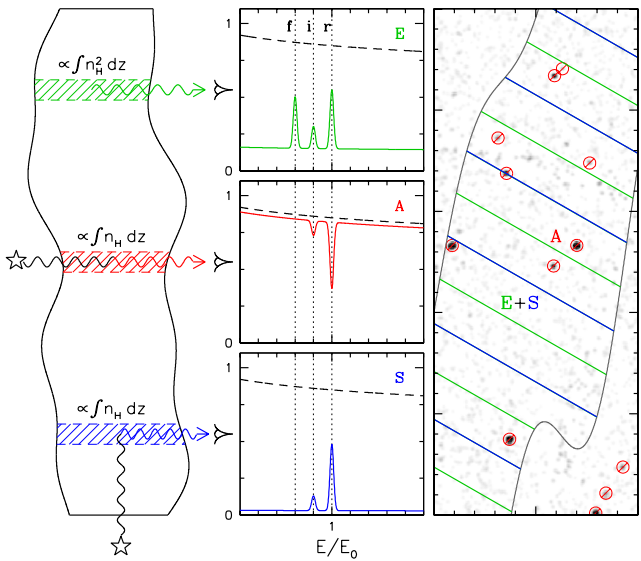}
\caption{{A schematic illustration of three main signatures of a WHIM layer in X-ray spectral band containing a triplet of lines from some helium-like ion (e.g. O VII). \textit{Left panel} depicts a WHIM layer illuminated by distant X-ray sources (shown as stars). The layer is seen (by a distant observer to the right) in intrinsic X-ray emission produced inside it (top), resonant absorption in the spectra bright background sources (middle), and resonant scattering of the isotropic CXB emission (bottom).} Amplitude of the first effect is proportional to the emission measure integrated over line-of-sight, for the latter two it is proportional to the integrated number density (i.e. column density) of the layer. \textit{Middle panel} demonstrates X-ray spectra characteristic for the the aforementioned signatures, with the prominence of the features strongly exaggerated for clarity in each case. Namely, the emission component (E) possesses both signification continuum and line emission, with comparable amplitude of the resonant(r), intercombination(i) and forbidden(f) triplet components. {The absorption spectrum (A) is the spectrum of the background source (black dashed line) with imposed resonant absorption lines at the redshift of the WHIM layer and also slight decrement in the continuum.} The scattered component (S) is heavily dominated by resonantly scattered emission lines of the most abundant ions, with no contribution from forbidden lines and very weak (Thomson scattered) continuum component. \textit{Right panel} illustrates the observing strategy for detection of the scattered component: emission of bright background sources (marked with crossed red circles), containing the absorption features induced by a WHIM filament (sky projection of which is shown as the hatched region), should be removed from the aperture, so that residual signal will contain both unresolved fraction of the CXB and emitted (E) plus scattered (S) radiation from the WHIM.}
\end{figure}

In fact, resonant scattering is not a true absorption process by itself, so that intensity decrement in the direction of the bright background sources is compensated by the intensity increment in all other directions (see Fig. \ref{fig:sketch}). Thus, the net effect would cancel out after integration over all directions in the case of isotropic incident radiation field, such as CXB, for instance. That is, a cloud that scatters isotropic radiation field is not visible neither as a bright nor a dark patch  on the sky.  However, a large fraction of CXB is composed of bright points sources, which can be resolved individually, and hence excluded at some level from any given aperture (as illustrated on the right panel in Fig. \ref{fig:sketch}). The remaining signal will then contain both the unresolved part of the background radiation (with similar absorption features as in resolved part, of course) plus resonantly scattered CXB, which is spatially extended, so that the excluded regions contribute only a small portion of overall signal from the filament. As a result, intrinsic thermal emission from the WHIM will be supplemented by resonantly scattered CXB, which might turn out to be of comparable amplitude, especially for regions of relatively low temperature and density, as has been pointed out by \citet{2001MNRAS.323...93C}. This effect not only significantly boosts possibility for WHIM detection in X-ray emission, but also changes the ratios of resonant lines to the continuum (i.e. their equivalent widths) and to forbidden lines, which is potentially an important diagnostic tool for physical conditions in this medium \citep{2001MNRAS.323...93C,2008SSRv..134..405P}.

{In this paper, we revisit the calculations of \citet{2001MNRAS.323...93C} and supplement them with calculations performed using the publicly-available photo-ionization code \texttt{Cloudy} \citep{2017RMxAA..53..385F} in order to produce a spectral model for the X-ray emission from the WHIM illuminated by CXB. We analyse basic characteristics of the predicted emission and compare it with the case when contribution of resonant scattering is ignored. The main results of \citet{2001MNRAS.323...93C} are confirmed and quantified in more detail over an extensive region of the density-temperature phase space relevant for the WHIM. We discuss the emissivity boost and corrections in diagnostics of WHIM based on emission-absorption comparison (e.g. \cite{2008SSRv..134..405P}) and emission line ratios induced by the contribution of the resonant scattering. We take advantage of the density-temperature distribution of the metal mass in the Local Universe extracted from a cosmological hydrodynamical simulation coupled with the ionization state calculations to highlight the parameter regions, for which these predictions are of the highest importance from the observational point of view. We illustrate the predicted X-ray emission from fiducial sheet-like (mean over-density $\sim5$) and filament-like (mean over-density $\sim30$) structures in the Local Universe in more detail, and briefly consider their detectability with the future X-ray missions. }


The structure of the paper reads as follows: in Section 2, we describe the calculations of the X-ray emission and present the resulting spectral model. In Section 3, we explore properties of this model and compare it to the purely thermal emission and emission from photoionized medium with no account for the resonant scattering. In Section 4, we discuss observational requirements and detectability of the predicted signal with current and future generations of X-ray observatories. We end up summarizing the main conclusions in Section 5.

\centerline{}

Throughout the paper, we adopt the base \textit{Planck} flat $\Lambda$CDM cosmology with the Hubble parameter $h=H_{0}/(100~\km~\s^{-1}~\Mpc^{-1})=0.678$ and the baryonic density parameter $\Omega_{b}h^2=0.0223$ \citep{2016A&A...594A..13P}. Consequently, the mean baryonic density of the Universe equals $ \left<\rho_b\right>=\Omega_{b}\,\frac{3H_0^2}{8\pi G}=4.2 \times 10^{-31} \g~\cm^{-3}$ in what follows ($G$ is the gravitational constant). This corresponds to the mean number density of hydrogen $ \left<n_H\right>=1.8\times10^{-7}$ cm$^{-3}~\left(\frac{X}{0.7}\right)$ at  $z\sim 0$, the hydrogen mass fraction  $X=0.7$ for solar metallicity and $X=0.75$ for primordial abundances. In what follows, we will adopt $n_0=2\times10^{-7}$ cm$^{-3}$ as a fiducial value for simplicity.

\section{Calculation}
\label{s:calculation}

\subsection{Physical conditions}
\label{ss:physcond}

~~~~~~~Let us consider a homogeneous isothermal slab of WHIM in the nearby Universe, i.e. at redshift $z\sim0$, that is characterized by hydrogen number density $n_{H}=(1-10^{3})\times n_0=(2\times10^{-7}-2\times10^{-4})$ cm$^{-3}$, temperature $T=(10^4-10^7)$ K\footnote{Formally, the gas with temperatures below $10^5$ K is not commonly regarded as WHIM, however, we will include it our consideration since it also turns out to be capable of producing significant emission in X-ray lines thanks to ubiquity of helium-like oxygen ions in regions of relatively low density.}, and metallicity $Z=(0.1-0.3)$, expressed in the units of the solar metallicity $Z_{\odot}$. 
Such a slab can be treated as an approximation for a part of bigger filament- or sheet-like structures, the regions with mean overdensities from few tens to several, and typical sizes $L$ of several Mpc to tens of Mpc, respectively. The scale at which non-linear evolution of the density field starts to develop corresponds to $R\sim8 h^{-1}\approx 12$ Mpc in the present-epoch Universe, so the space density of structures with $L\gtrsim R$ should fall exponentially. 

The aforementioned parameters imply Thomson optical depth at level

\begin{equation}
\tau_T\approx 5\times 10^{-6}\left(\frac{n_{H}}{n_0}\right)~\left(\frac{L}{10\rm Mpc}\right), 
\label{eq:taut}
\end{equation}
where the ratio of electron and hydrogen number densities $x_e=n_e/n_H=1.21$ was assumed.

Evidently, the optical depth given by Eq.(\ref{eq:taut}) is sufficiently small both for sheets and filaments, so they are effectively transparent for the bulk of cosmic background radiation, especially for the cosmic X-ray background radiation produced by collective emission of AGN population in the Universe. Due to the very low density of the intergalactic medium, the background radiation field has a very prominent effect on thermal and ionization state of this gas \citep[e.g.][]{2009MNRAS.399..574W}. Also, time-scales of reaching thermal equilibrium turn out be long compared to the dynamic time-scales in many cases, so that non-equilibrium states should be allowed for when considering WHIM \citep[e.g.][]{2006PASJ...58..641Y}. This motivates the detailed ionization balance modelling over the whole extent of the density-temperature diagram relevant for the WHIM \citep[see][for examples of such calculations coupled with cosmological hydrodynamic simulations]{2003PASJ...55..879Y,2010MNRAS.407..544B,2011ApJ...731....6S,2018MNRAS.476.4629P}. 

\subsection{Cosmic radiation field}
\label{ss:radiation}

\begin{figure}
\centering
\includegraphics[bb=20 200 570 680,width=1.\columnwidth]{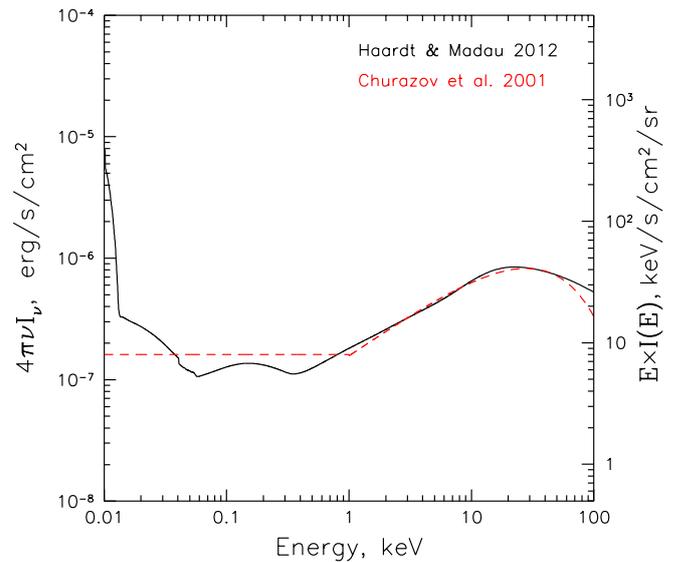}
\caption{Intensity and spectral shape of the cosmic UV/X-ray background radiation field. The solid black line shows the \citealt{2012ApJ...746..125H} model at $z=0$, the dashed red line is the phenomenological approximation to the measured CXB as used in \citealt{2001MNRAS.323...93C}.}
\label{fig:inspec}
\end{figure}

The cosmic UV/X-ray background radiation field and its redshift evolution is very well studied \citep{2012ApJ...746..125H}. Inasmuch we are interested primarily in WHIM at $z\sim0$, a simple approximation for the observed CXB intensity and spectral shape can been used:
\begin{equation}
\label{eq:cxb}
I(E)=I_0\,\left\{
\begin{array}{ll}
 E^{\,-2} \times ~~e^{-\frac{1}{E_c}}, &  E \le 1\,{\rm keV} \\
 E^{\,-1.3}\times  e^{-\frac{E}{E_c}}, &  E \ge 1\,{\rm keV},
\end{array} 
\right.
\end{equation}
where  $I_0\approx8~{\rm phot\,cm^{-2}\,s^{-1}\,keV^{-1}\,sr^{-1}}$, $E_c=40\,{\rm keV}$ \citep[e.g.][]{1992ARA&A..30..429F,1998A&A...334L..13M}. It is this approximation that has been used for calculations in \citet{2001MNRAS.323...93C}, and despite slight differences in the X-ray part and absence of the bump below 13.6 eV (see Fig. \ref{fig:inspec}), usage of it results in almost identical ionization balance compared to the calculations with the \cite{2012ApJ...746..125H} radiation field, as we confirmed by calculations with both \texttt{Cloudy} and our own photoionization code (as described next).

\subsection{Ionization state}
\label{ss:ionstate}

~~~~~For consistency with \citet{2001MNRAS.323...93C}, we use the same recipes when calculating the ionization equilibrium and the line emission\footnote{Although more recent compilations of atomic data do exist, we have verified that, for the problem at hand, the changes in the ionization equilibrium and the final line fluxes are insignificant.}. Namely, photoionization cross sections, rates of collisional ionization and photo and dielectronic recombination used to calculate ionization equilibrium are those of \cite{1995A&AS..109..125V,1996ApJS..103..467V}. Ejection of multiple electrons, which may follow inner shell ionization, has been switched off. The medium is considered to be optically thin for both the incident and scattered/emitted radiation field. In this limit, and under assumption of ionization equilibrium (but not necessarily in thermal balance, viz., cooling=heating), the ionization state of the gas is fully defined by its number density and temperature. Because of that we perform calculations on an extensive density-temperature grid that covers the parameter space region relevant for WHIM.

\begin{figure*}
\centering
\includegraphics[bb=50 180 600 700,width=0.32\textwidth]{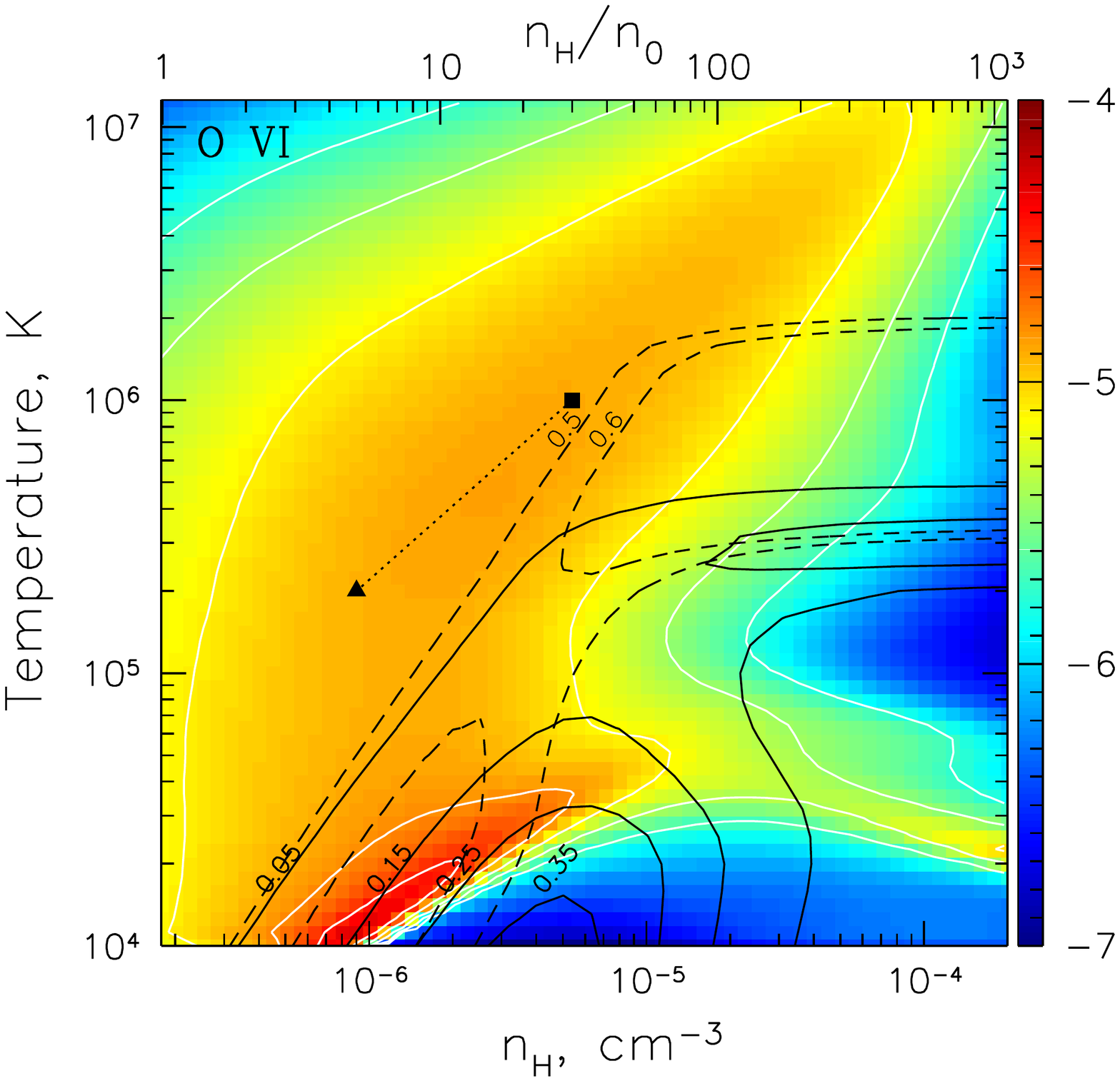}
\includegraphics[bb=50 180 600 700,width=0.32\textwidth]{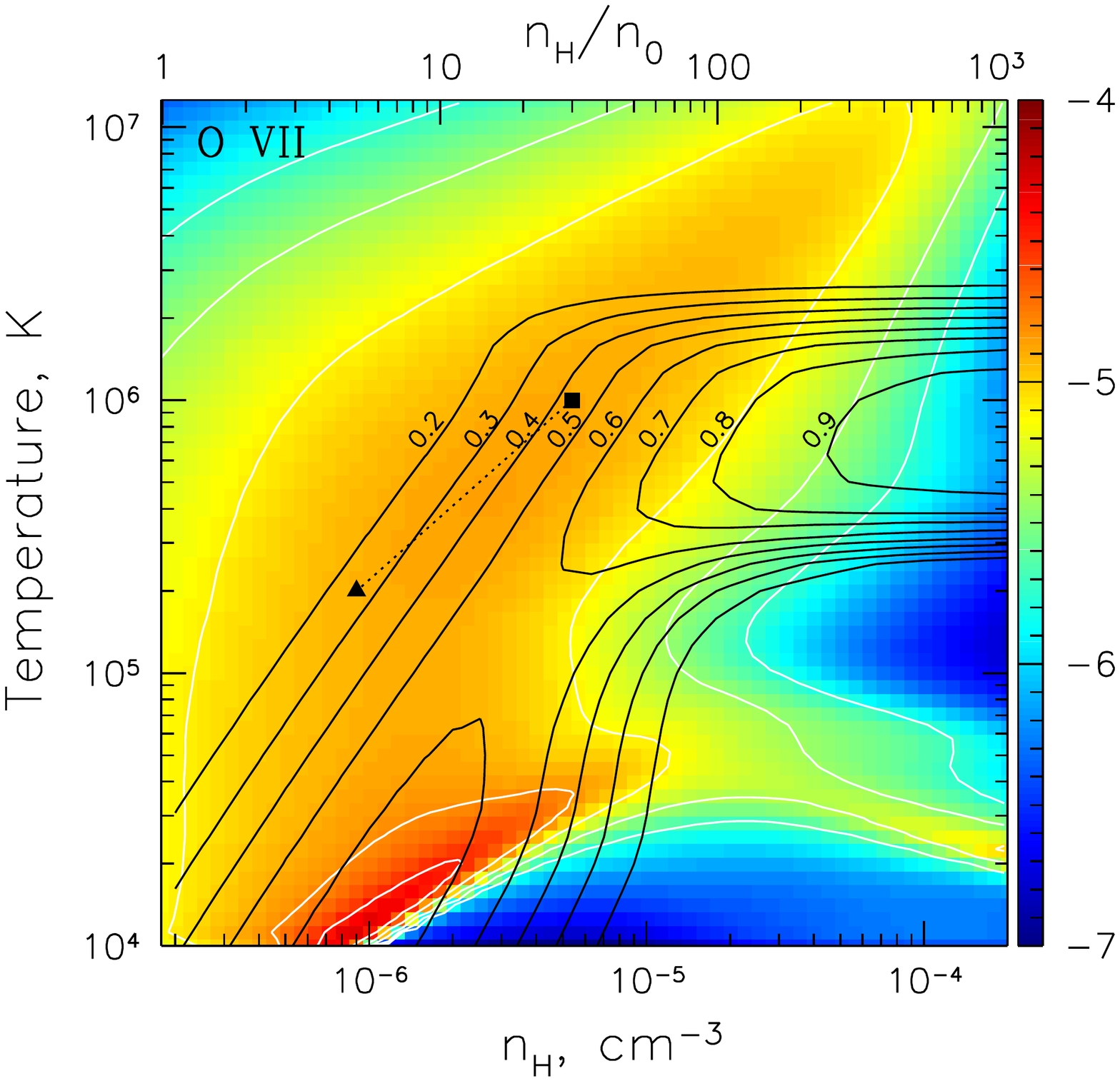}
\includegraphics[bb=50 180 600 700,width=0.32\textwidth]{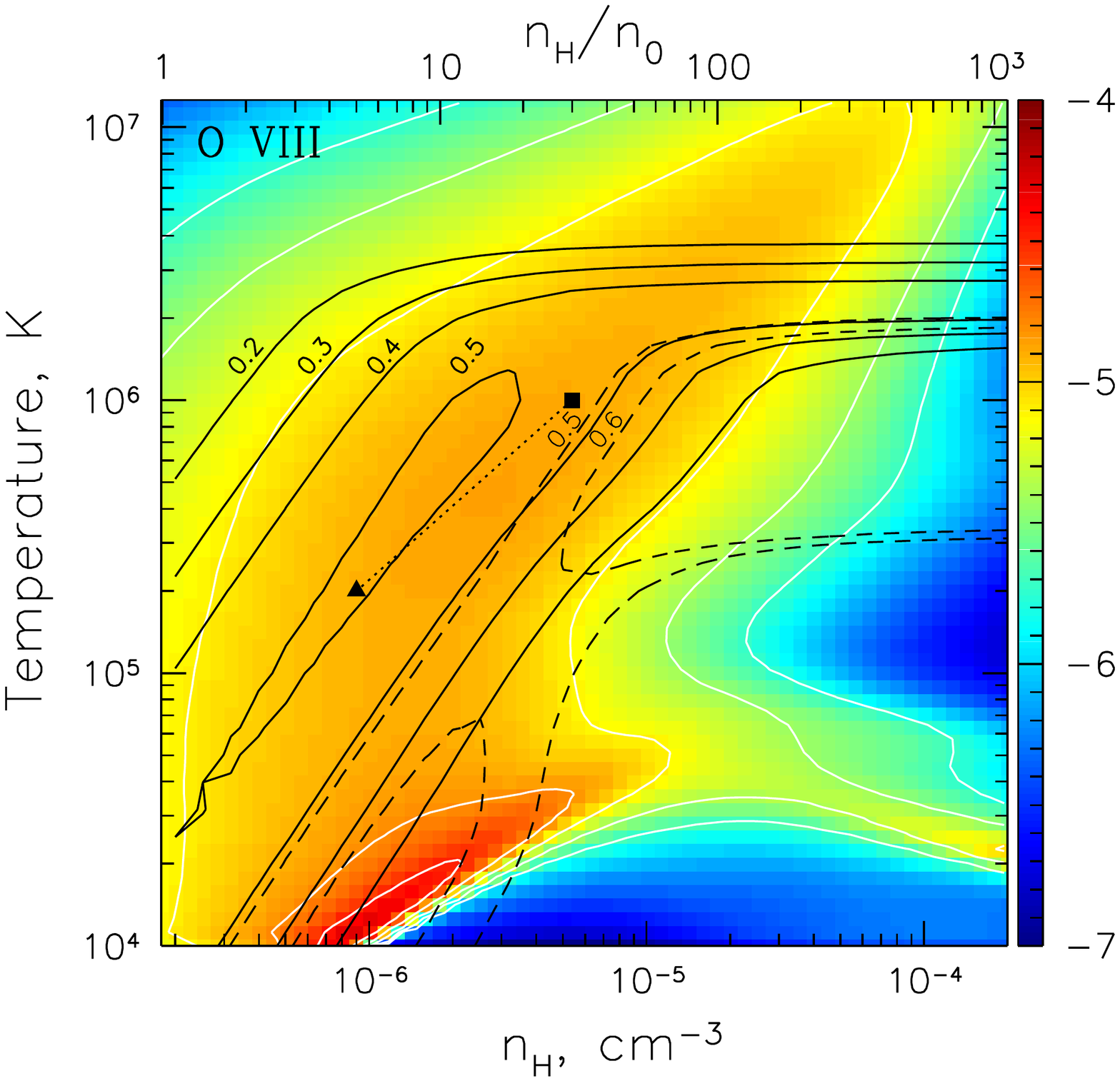}
\caption{Differential distribution of mass $\frac{dM(n_H,T)}{d\log n_H\,d\log T}$, extracted from a $z\sim0$
snapshot of the \textit{Magneticum} cosmological hydro-simulation \citep{2016MNRAS.463.1797D}, color-coded on a logarithmic scale (arbitrary units) with corresponding white contours spaced by a factor of two. Black solid contours depict O VI (left), O VII (middle) and O VIII (right) ionization fractions. The dashed contours in the left and right panels indicate O VII ionization fraction (only 0.5 and 0.6 levels are shown). The black solid triangle and square connected by a dotted line mark the parameters of sheet-like and filament-like structures illustrated in the text in more detail.}
\label{fig:dti_o_mass}
\end{figure*}

\begin{figure*}
\centering
\includegraphics[bb=50 180 600 700,width=0.32\textwidth]{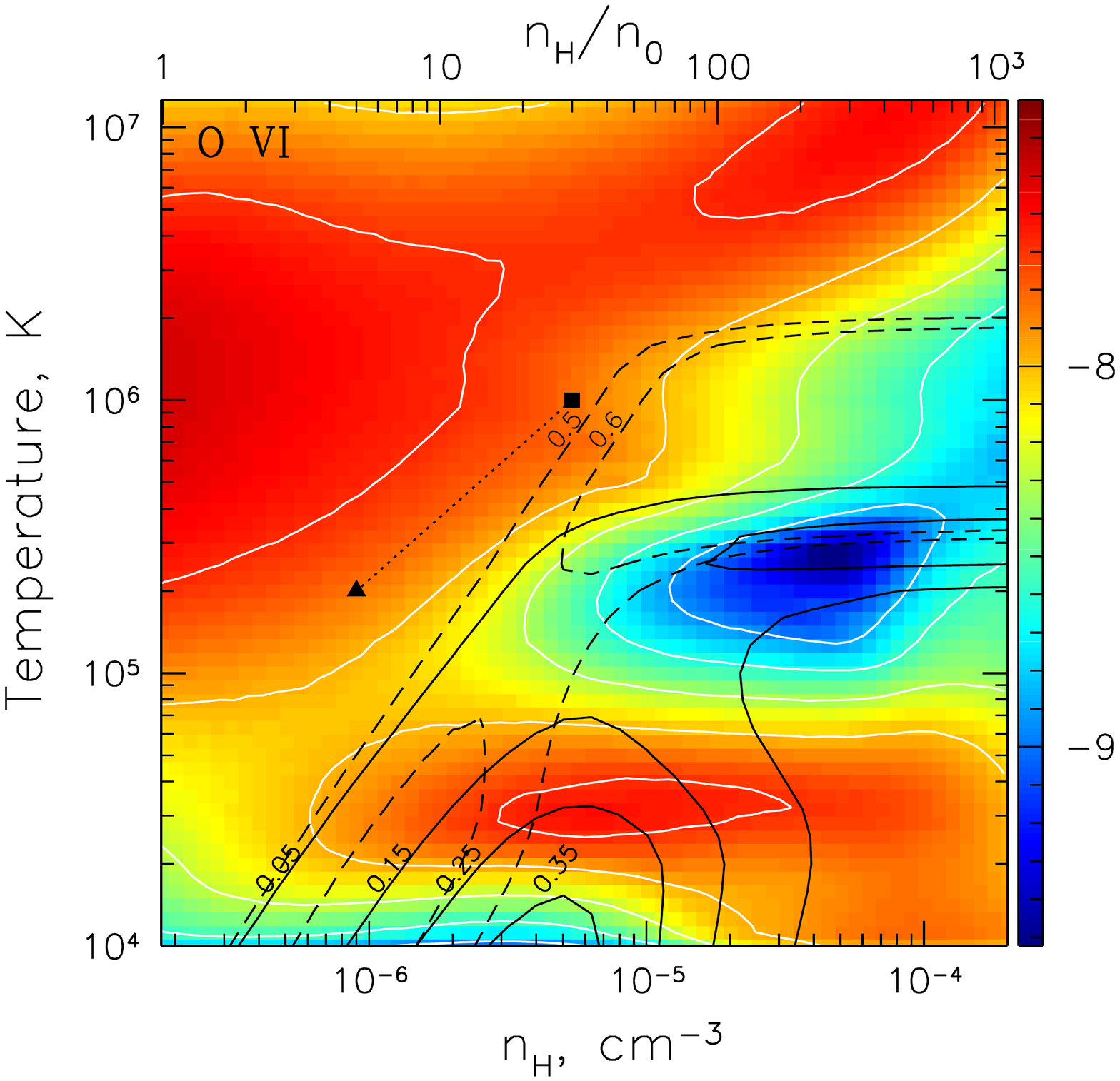}
\includegraphics[bb=50 180 600 700,width=0.32\textwidth]{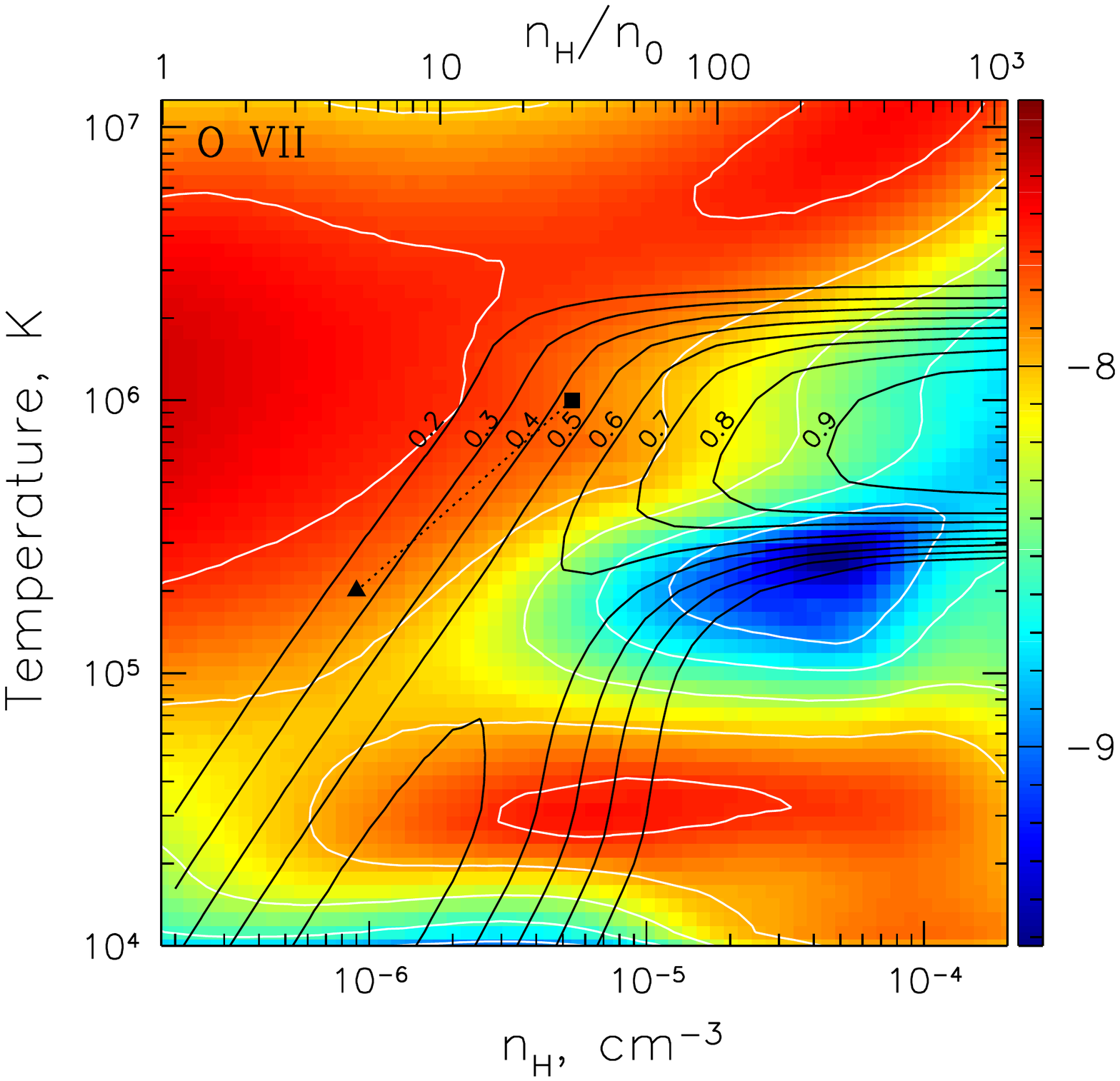}
\includegraphics[bb=50 180 600 700,width=0.32\textwidth]{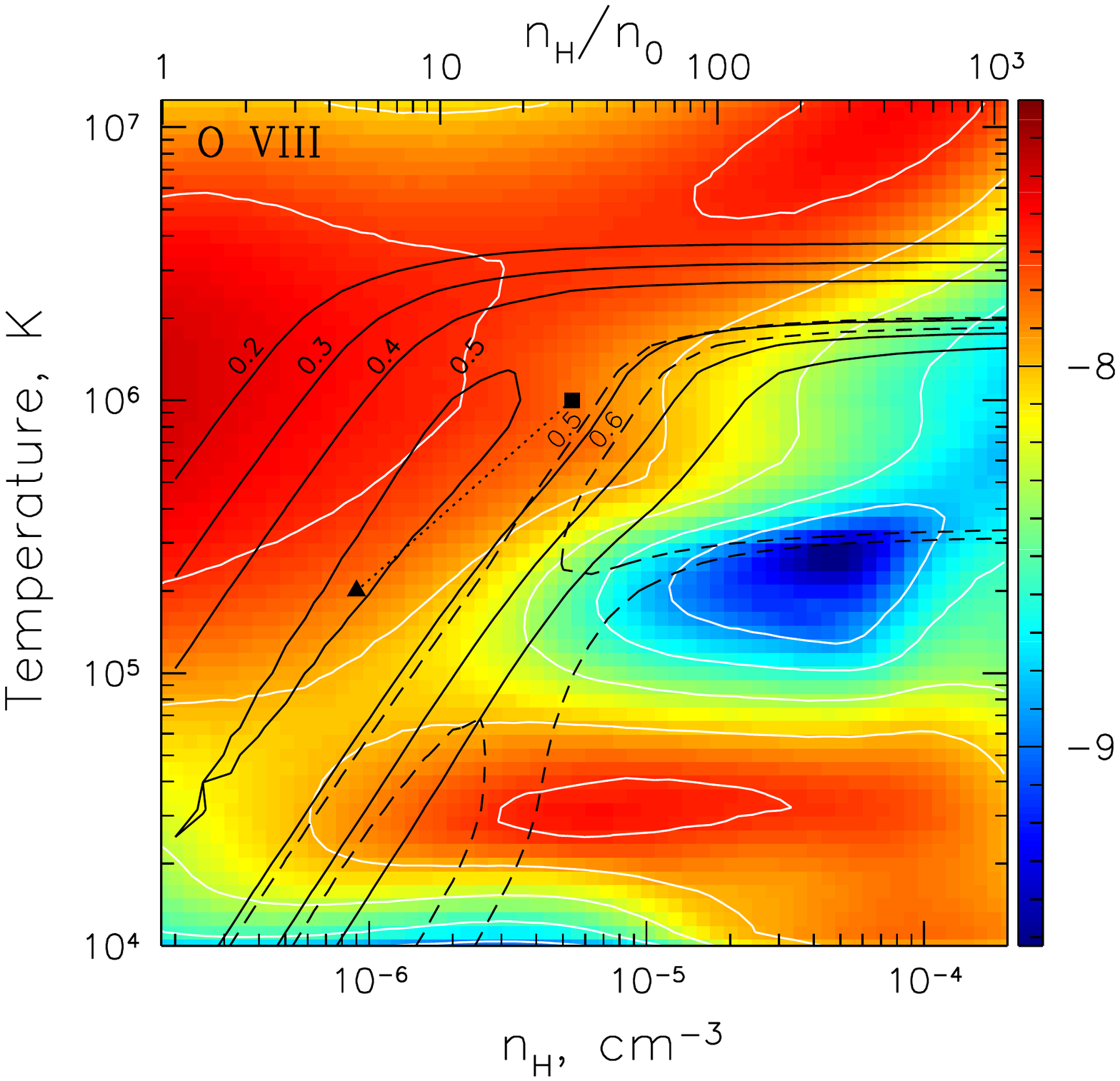}
\caption{Differential distribution of metal mass $\frac{Z\times dM(n_H,T)}{d\log n_H\,d\log T}$, extracted from a $z\sim$ snapshot of the \textit{Magneticum} cosmological hydro-simulation \citep{2016MNRAS.463.1797D}, color-coded on a logarithmic scale (arbitrary units) with corresponding white contours spaced by a factor of two. Black solid contours depict O VI (left), O VII (middle) and O VIII (right) ionization fractions. The dashed contours in the left and right panels indicate O VII ionization fraction (only 0.5 and 0.6 levels are shown). The black solid triangle and square connected by a dotted line are same as in previous plot.}
\label{fig:dti_o_massmet}
\end{figure*}

We perform an identical set of calculations using \texttt{Cloudy}\footnote{Version 17.00, last described in \citet{2017RMxAA..53..385F}.} (in the optically thin limit as well) with  the radiation field given by Eq.(\ref{eq:cxb}) and also the one of \citet{2012ApJ...746..125H}. All three sets agree well, and in Figures \ref{fig:dti_o_mass} and \ref{fig:dti_o_massmet} we show ionization fractions of hydrogen-, helium-, and lithium-like oxygen (O VIII, OVII and OVI hereafter) resulting from the latest set (i.e. \texttt{Cloudy} with the \citet{2012ApJ...746..125H} radiation field)\footnote{{This consistency can be seen by comparing the ionization fraction contours in Figure \ref{fig:dti_o_mass} with those in the Figure 3 of \citet{2001MNRAS.323...93C}.}}.   

{
In order to illustrate the mass and metallicity distribution of the WHIM on the same plots, we are using a $z\sim0$ data extraction from the recent, cosmological, hydrodynamical simulation {\it Box2/hr} of the {\it Magneticum}\footnote{\href{www.magneticum.org}{www.magneticum.org}} simulation set. More details on the simulation can be found in \citet{2014MNRAS.442.2304H,2016MNRAS.463.1797D}. Most importantly for the purpose of our study, the simulation follows the detailed evolution of various metal species -- and their relative composition -- in the IGM and ICM due to continuous enrichment by supernovae of type Ia and type II (SNIa and SNII), and asymptotic giant branch (AGB) star winds. Those are predicted self-consistently based on the underlying evolution of the stellar population. Following these physical processes allows such simulations to reproduce the detailed metallicity distribution within the ICM \citep[see][]{2017Galax...5...35D,2017MNRAS.468..531B}. Even more importantly, in combination with the AGN feedback, this allows to achieve significant enrichment of the WHIM already at very high redshifts (e.g. $z\approx2-3$), see \citet{2017MNRAS.468..531B}. For simplicity, the total metallicity predicted by the simulations is used in what follows.
}

\begin{figure*}
\centering
\includegraphics[bb=50 180 620 700,width=0.32\textwidth]{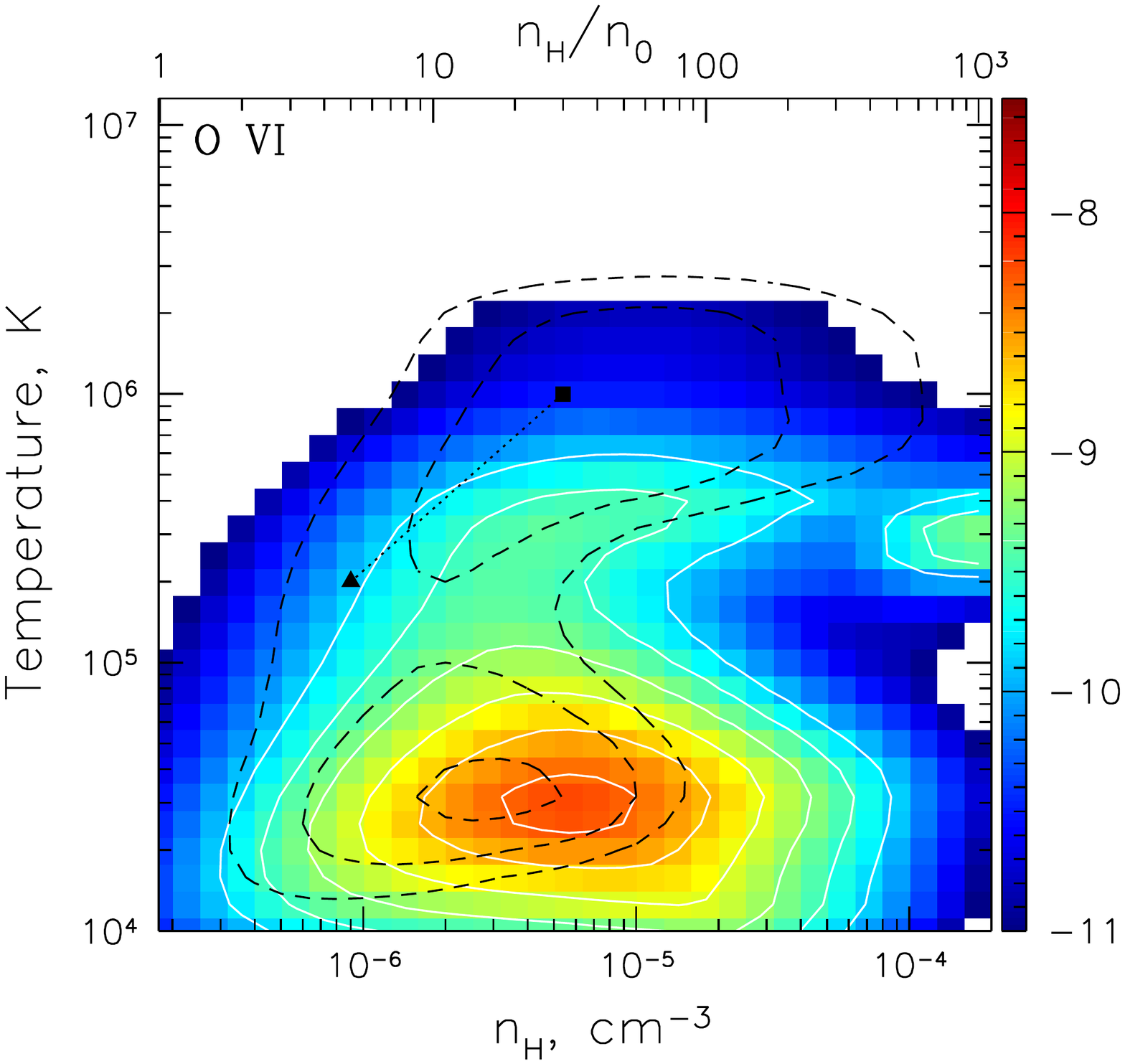}
\includegraphics[bb=50 180 620 700,width=0.32\textwidth]{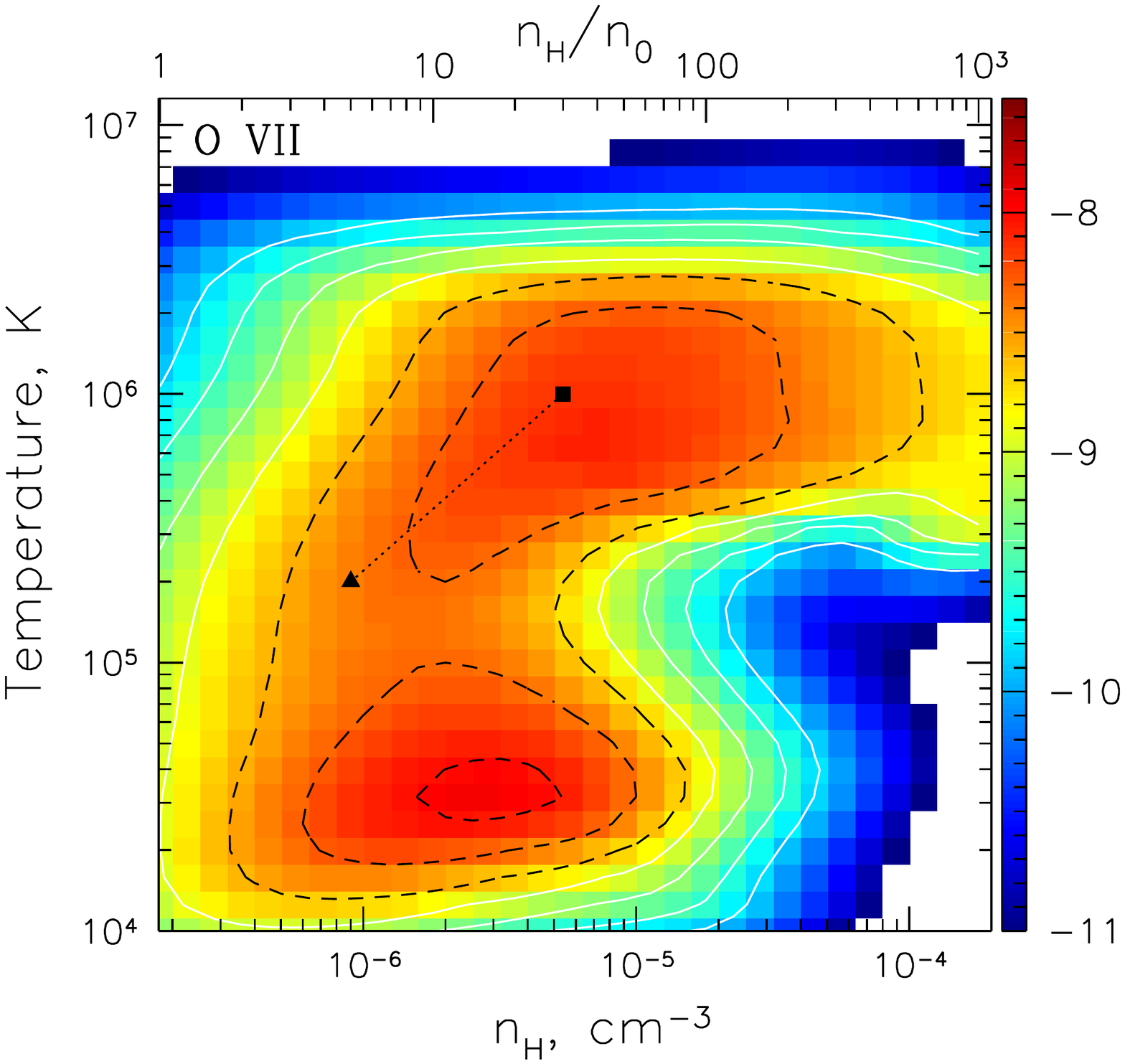}
\includegraphics[bb=50 180 620 700,width=0.32\textwidth]{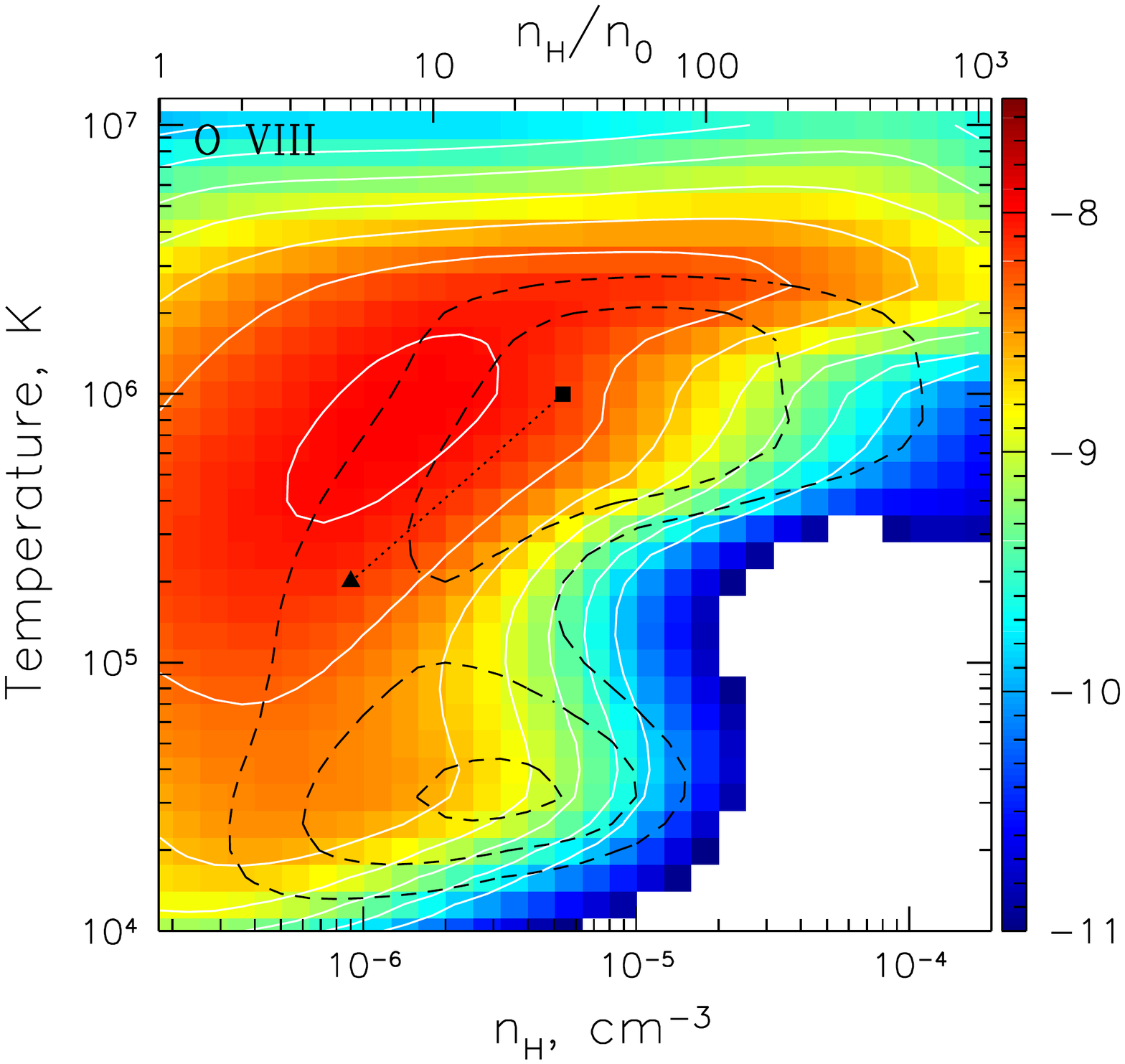}
\caption{Differential distribution of a particular ion mass $\frac{x_{i}\times Z\times dM(n_H,T)}{d\log n_H\,d\log T}$ (O VI - left, O VII - middle, and O VIII - right panel), based on $z\sim$ distributions from \textit{Magneticum} cosmological hydro-simulation \citep{2016MNRAS.463.1797D} and ionisation balance calculation described in Section \ref{ss:ionstate}. It is color-coded on a logarithmic scale (arbitrary units) with corresponding white contours spaced by a factor of two. The dashed contours in the left and right panels show the three highest contours for O VII from the middle panel. The black solid triangle and square connected by a dotted line are same as in previous plots.}
\label{fig:dti_o_massmetion}
\end{figure*}

Evidently, O VII traces both high ($T\approx 10^6$ K) and low ($T\approx 3\times 10^4$ K) temperature portions of the metal-enriched WHIM, while O VIII is present mostly in the hotter part, and O VI - in the colder one (see Fig.\ref{fig:dti_o_massmetion}). 

It is worth mentioning, that the photo-ionization time-scale for oxygen atoms in the CXB radiation field $t_{ion}=\left[4\pi\int I(E)\sigma_{ph}(E)dE\right]^{-1}$, $\sigma_{ph}(E)$ is the relevant photoionization cross-section, equals $\sim 10^{10}$ yrs, $\sim 3\times10^{9}$ yrs and $\sim 3\times10^{8}$ yrs for O VIII, OVII and OVI respectively, i.e. shorter than the Hubble time, $t_{H}=H_{0}^{-1}\approx 1.4\times 10^{10}$ yrs. However, the recombination time-scale turns out to be longer than the Hubble time for gas with over-densities $\delta \lesssim 10$ and temperatures $T\gtrsim 3\times 10^5$ K (see, e.g., Fig. 3 in \citealt{2006PASJ...58..641Y}). That means regions falling in the left upper corner on the density-temperature diagram (specifically, Fig. \ref{fig:dti_o_massmetion}), might be considerably over-ionized compared to the equilibrium estimates, hence favouring O VIII ions over O VII ones, in particular.  

\subsection{Emission and scattering by WHIM}
\label{ss:emission}

~~~Intrinsic volume emissivity [erg s$^{-1}$ cm$^{-3}$] of WHIM in a spectral window of width $\delta E$ centred on $E_{0}$ is generally expressed as
\begin{equation}
\epsilon_{em}(E_0,\delta E)=n_{H}n_{e}~\Lambda_{T,n_{H}}(E_0,\delta E)
\label{eq:emm_em}
\end{equation}
with $ \Lambda_{T,n_{H}}(E,\delta E)$ [erg s$^{-1}$ cm$^{3}$] standing for the plasma emissivity function integrated over the specified spectral window, given the temperature $T$ and number density $n_{H}$ of the gas (in the collisional equilibrium limit, $ \Lambda$ is a function of temperature only, while in the low-density photo-ionized matter, population of the levels is also affected by the recombination dynamics, giving rise to the $\Lambda$ dependence on $n_{H}$ as well).

The photon volume emissivity (photons s${^{-1}}$ cm$^{-3}$) of radiation resonantly scattered in the line with central energy $ E_{\ell}$ is given by
\begin{equation}
\epsilon_{rsc}(E_{\ell})=4\pi I(E_{\ell})\,n_i\sigma_{0},
\label{eq:emm_sc}
\end{equation}
$\sigma_{0}=(\pi e^2/m_e c)~(h/keV)~f_{ul}$ cm$^2$ keV, $f_{ul}$ is oscillator strength of the corresponding $u \rightarrow l$ transition, $e$ and $m_e$ are electron's charge and mass, $h$ is the Planck's constant, $keV=1.60219\times 10^{-9}$ erg. For the resonant scattering off the strongest transitions (having $f_{ul}=0.4-0.7$) in helium- and hydrogen-like oxygen ions, one has $\sigma_0/\sigma_T\sim10^5$ keV, and it is approximately the same along these isoelectronic sequences.

The number density of ions of interest is expressed as
\begin{equation}
n_{i}=n_{H}\,\left(\frac{n_{Z}}{n_{H}}\right)\,x_{i},
\label{eq:tau_res}
\end{equation}
where $(n_{Z}/n_{H})$ is the relative abundance of particular element $ Z $ with respect to hydrogen as determined by some abundance set (e.g. the solar one), and $ x_{i}$ stands for the fraction of the particular ionization state.

Equivalent width of the resonantly scattered lines with respect to the Thomson scattered continuum equals
\begin{equation}
EW_{s}=\frac{n_{i}\sigma_{0}}{n_{e}\sigma_{T}}=\pi\frac{e^2}{m_ec^2}\frac{hc}{\sigma_T}~f_{ul}~\frac{n_{Z}}{n_{H}}~\frac{x_i}{x_e}.
\label{eq:ews}
\end{equation} 
Hence, for gas metallicity $\sim 0.3 Z_{\odot}$, i.e. $n_Z/n_H \sim 10^{-4}$ and $x_i/x_e\sim 0.5$, one might expect $ EW_s\sim 15$ keV. That means the corresponding line emission should be $\sim30$ times larger than the total continuum flux between 0.5 and 1 keV, the spectral band where this line is located (see also Table 1 in \mbox{\citealt{2001MNRAS.323...93C}}, where equivalent widths of the other prominent lines are listed for the solar abundance of elements).

The equivalent width of the corresponding absorption lines in the spectrum bright background sources can be readily expressed as 
\begin{equation}
EW_{a}=\tau_T~EW_{s},
\end{equation}
so combining Eq.(\ref{eq:taut}) and Eq.(\ref{eq:ews}), one might expect $ EW_{a}\sim 0.1$ eV at $E_{\ell}\sim0.5$ keV (i.e. $\sim 5$ mA at $\lambda_{\ell}\sim$ 25 A), given that $\tau_T\sim 10^{-5}$. This is a well known result from the searches of WHIM using absorption features in spectra of bright background sources \citep[see, e.g.,][]{2018Natur.558..406N}.

The optical depth in the line center is estimated as
\begin{equation}
\tau_{0}\sim\tau_{T}~\frac{EW_s}{E_{\ell}}~\frac{E_{\ell}}{\Delta E},
\end{equation}
where $\Delta E$ is the characteristic broadening of the line, so that $\Delta E/E_{\ell}\sim\Delta z\sim L H_{0}/c\sim 3\times10^{-3}h~(L/10 \mathrm{Mpc})\sim 2\times10^{-3}~(L/10 \mathrm{Mpc})$ (thermal broadening equals $ \sim 2\times 10^{-4} \sqrt{T/10^{6} \mathrm{K}}$, i.e. $\sim 0.1$ eV at 0.5 keV, for oxygen ions, so it can be neglected in most cases). Hence, for a line at $E_{\ell}\sim0.5$ keV with $ EW_s\sim 15$ keV, we get $ \tau_{0}\sim 1.5\times 10^{4}\tau_{T} $, that basically validates usage of the optically thin limit for structures with $ \tau_{T}\lesssim 10^{-4}$, which are indeed the primary object of exploration here. 

%

Comparing Eq.(\ref{eq:emm_sc}), Eq.(\ref{eq:ews})  and Eq.(\ref{eq:emm_em}), one can see that
\begin{equation}
\mathcal{R}(E_{\ell},E,\delta E)=\frac{E_{\ell}\epsilon_{rsc}(E_{\ell})}{\epsilon_{em}(E,\delta E)}=
\frac{4\pi\sigma_{T}~E_{\ell}~I(E_{\ell})~EW_s(E_{\ell})}{n_{H}~\Lambda_{T,n_{H}}(E,\delta E)}.
\end{equation} 

Numerically, this can be expressed as  
\begin{equation}
\begin{array}{c}
\bigskip
\mathcal{R}(E_{\ell},E,\delta E)\simeq 1.0\times \left(\frac{E_{\ell}}{0.5~\mathrm{keV}}\right)
\left(\frac{I(E_{\ell})}{20~\mathrm{ph~s^{-1}~keV^{-1}~sr^{-1}}}\right)\times\\
\times\left(\frac{EW_s}{15~\mathrm{keV}}\right)
\left(\frac{10^{-23}~\mathrm{erg~s^{-1}~cm^{3}}}{\Lambda_{T,n_{H}}}\right)
\left(\frac{2\times 10^{-7}~\mathrm{cm^{-3}}}{n_{H}}\right).
\end{array}
\end{equation} 

The characteristic value $10^{-23}~\mathrm{erg~s^{-1}~cm^{3}}$ used above for $\Lambda_{T,n_{H}}$  corresponds actually to the plasma emissivity integrated over the whole energy range, i.e. to the radiative cooling function, for a gas with temperatures  $10^4-10^{7}$ K and $\sim 0.1 Z_{\odot}$ metallicity \citep[see, e.g.,][]{2007ApJS..168..213G,2017ApJS..228...11G}. For $E\gg \kappa T$ and $\delta E \sim E$, $\Lambda_{T,n_{H}}(E,\delta E)$ is likely $\sim e^{E/\kappa T}$ times smaller than {the total (integrated over all energies) one}. Clearly, this factor becomes large for $E=0.5-1$ keV and $T\lesssim 10^{6}$ K, simply reflecting the fact that gas in collisional equilibrium at these temperatures barely emits above 0.3 keV. Given that the highly-ionized atoms of oxygen do survive across the wide range of WHIM temperatures, a very strong boost of the X-ray emission from the colder phases might be supplied by resonant scattering compared to the intrinsic thermal emission of the WHIM. 

Intrinsic emission of the low-density and low-temperature WHIM in lines of highly ionized metals is powered mostly by recombination at upper levels of corresponding transitions, in contrast to the collisionally-dominated excitation at higher densities and temperatures. In ionization equilibrium, recombination rate equals direct ionization rate (in a simple 'two-stage' approximation), so the intrinsic line emissivity can be estimated as
\begin{equation}
\epsilon_{em}(E_\ell)\sim g(E_\ell)~n_{i}\int_{\chi} 4\pi I(E)\sigma_{ph}(E)dE, 
\label{eq:emm_lin}
\end{equation}   
where $g(E_\ell)<1$ stands for the branching ratio for the specific line of interest and $\chi>E_\ell$ is the photo-ionization threshold energy. For a sufficiently smooth $I(E)$, the photo-ionization integral can be expressed in terms of the cross-section at the absorption edge $\int_{\chi}I(E)\sigma_{ph}(E)dE=I(\chi)\sigma_{ph}(\chi)\Delta E_{ph}$, with $\Delta E_{ph}\simeq \chi/4.$ for $ I(E)\propto E^{-2}$, as is approximately the case for CXB below 1 keV (see Section \ref{ss:radiation}), and taking into account $ \sigma_{ph}(E)=\sigma_{ph}(\chi)(E/\chi)^{-3}$ shape of the cross-section above the threshold. The threshold cross-section for inner shell electrons is $\sigma_{ph}(\chi)=A\pi (a_0/Z_i)^2 N_e$, where $ a_0=5.3\times10^{-9}$ cm is the Bohr radius, $Z_i$ effective charge of the nuclei, $N_e$ the number of inner shell electrons, and $A$ is the amplitude factor. For He-like oxygen, $Z_i=7$, $ N_e=2$ and $A=0.067$, so $\sigma_{ph}(\chi)=2.4\times 10^{-19} $ cm$^2$. 

The ratio of scattered to intrinsic line emission can then be expressed as
\begin{equation}
\mathcal{R}(E_{\ell})\sim\frac{\epsilon_{rsc}(E_\ell)}{\epsilon_{em}(E_\ell)}\sim\frac{4\sigma_{0}}{g(E_\ell)\sigma_{ph}(\chi)\chi},
\label{eq:scemtalin}
\end{equation} 
where $I(\chi)\approx I(E_{\ell})$ for $ E_\ell\simeq \chi$ was assumed. Thus, this ratio is determined essentially by the difference in the effective photo-ionization and resonant scattering cross-sections, and the corresponding branching ratio $ g(E_{\ell})$ for a specific line, with no dependence on physical conditions in the gas at all (including metallicity), as far the adopted simplified assumptions are valid. For He-like oxygen, $\chi$=0.74 keV, while $ \sigma_0/\sigma_T\simeq 10^5$ keV and $g(E_\ell)\sim 0.1$ for the strongest resonant line. As a result, one has $ \mathcal{R}(\mathrm{OVIIr})\simeq 15$ (actually, for CXB-like spectrum $I(\chi)\approx 0.6\,I(\mathrm{OVIIr})$, so more accurate estimate is $ \mathcal{R}(\mathrm{OVIIr})\simeq 15/0.6=25$). 

For the intercombination transitions in the He-like triplets, this ratio is a factor of $10$ smaller (because $f_{ul}$ is smaller by this factor), and it equals zero for the forbidden components. As a result, the resonant scattering not only impact the total intensity of X-ray emission from WHIM, but also the relative amplitude of the resonant lines compared to the continuum and non-resonant lines. As a result, these effects should be taken into account when considering corresponding line diagnostics for the physical conditions in WHIM. In Section \ref{s:results} we quantify these effects in more detail based on the spectral calculations described below. 
  
\subsection{Spectrum calculation}
\label{ss:calculation}

~~~In our code, thermal emission from the WHIM is calculated by means of the MEKA \citep{1985A&AS...62..197M,1986A&AS...65..511M,1993A&AS...97..443K} code, as implemented in the software package XSPEC v10 \citep{1996ASPC..101...17A}, with the ionization fractions for each element supplied by the ionization balance calculations described above. The intensity of the scattered emission is calculated from the optical depths derived using  the line parameters (wavelengths, oscillator strengths, etc.) from the list of the most important resonant lines compiled by Dima Verner (http://www.pa.uky.edu/$\sim$verner/atom.html). For both calculations, the input metal abundances are specified by the single metallicity factor $Z$ with respect to the solar abundances from \citet{1992PhyS...46..202F}. The resulting line optical depths are consistent with the one calculated with \texttt{Cloudy} using the same input metal abundance set (see Fig. \ref{fig:odsp_sh} and Fig. \ref{fig:odsp_fi}). 

In \texttt{Cloudy}, emission and scattering from a WHIM `slab' are produced self-consistently with the ionization balance calculation. However, there are a number of subtleties in the process of saving and interpreting the corresponding spectral predictions, which require to be treated with care.

Namely, in \texttt{Cloudy} terminology, the case considered here has `open geometry', and we are interested in `reflected' emission, i.e. excluding attenuated emission of the illuminating source, viz. CXB.
In such situation, the scattering attenuation of the incident radiation field is treated according to the \citealt{1905ApJ....21....1S} prescription, i.e. the attenuation factor is $f=(1+0.5\tau_s)^{-1}$, not $f=\exp(-\tau_s)$ (see section 10.8 in the \texttt{Cloudy}'s manual \textit{Hazy 1}). In the limit of small optical depth $\tau_s\ll1$, the scattering decrement $(1-f)$ then equals $0.5\tau_s$. This approach has to be modified, however, if the bulk of the incident radiation field is composed of point sources, which can be resolved individually and excluded from the aperture. The spectrum of such sources is attenuated by a factor of $f=\exp(-\tau_s)$, of course, while the diffuse scattered emission is $ \propto (1-f)\approx\tau_s  $ for $\tau_s\ll1$. According to this, the contribution of scattered emission to the observed emission from a WHIM `slab' turns out to be factor of 2 lower in \texttt{Cloudy} predictions compared to our calculations.

Since resonant scattering in \texttt{Cloudy} is treated as line pumping by continuum emission, it can be artificially turned off by the `no induced processes' command. For the situation in hand, usage of this command does not affect any other important aspect of the simulation, but allows us to separate intrinsic emission from the resonantly scattered one, and by doing this verify that the latter turns out to be a factor of two smaller compared to predictions based solely on the corresponding optical depths. With that in mind, we have combined \texttt{Cloudy} calculations with and without 'no induced processes' command, so that in the end three spectral models have been produced for the intrinsic emission, resonantly scattered emission, and the total emission, which is the sum of former two. After doing this, the resulting predictions appear to be fairly consistent with the predictions of our code, both for the intrinsic and scattered components. 

{Additionally, we have checked that the thermal emission predicted by our code using MEKA is consistent with the \texttt{Cloudy} predictions and with the predictions of the APEC model \citep{2012ApJ...756..128F}.  In the spectral band of interest here, viz. $\sim$0.3-1.2 keV, noticeable differences are present only for weak lines and, sometimes, relative normalization of the continuum, which constitutes the far exponential tail of the bulk thermal emission in this case. Besides that, when comparing predictions of various codes and table spectral models, one should be aware of the inaccuracies induced by interpolation of the spectra calculated on some parameters grid. Clearly, this is relevant mostly for the predicted thermal emission and its contribution to the total intrinsic emission from the photo-ionized gas, which scales as square of the number density and which is highly sensitive to the gas temperature via strong dependence of the collisional excitation rate on it. In the situation when the gas ionization state and emission is governed mostly by the external radiation field, as is the case for the bulk of the WHIM considered here, interpolation-induced inaccuracies should not be a major issue.}

\section{Results}
\label{s:results}

~~~~~We illustrate predictions of the calculated spectral models on two examples considered also in \citet{2001MNRAS.323...93C}: a sheet-like structure with $n_H=10^{-6}$ cm$^{-3}$, $T=2\times10^{5}$ K and $\tau_{T}=2\times10^{-5}$, and a filament-like structure with $n_H=6\times10^{-6}$ cm$^{-3}$, $T=10^{6}$ K and $\tau_{T}=10^{-4}$. For simplicity, we take $Z=0.3$ in both cases. Many characteristics of the predicted emission, however, do not depend strongly on $Z$ as far as emission lines of the  most abundant ions dominate over the continuum emission. The absolute amplitude of the WHIM emission (both intrinsic and scattered) is, of course, approximately linearly proportional to $Z$ in this case.

\begin{figure*}
\centering
\includegraphics[bb=50 180 600 700,width=0.31\textwidth]{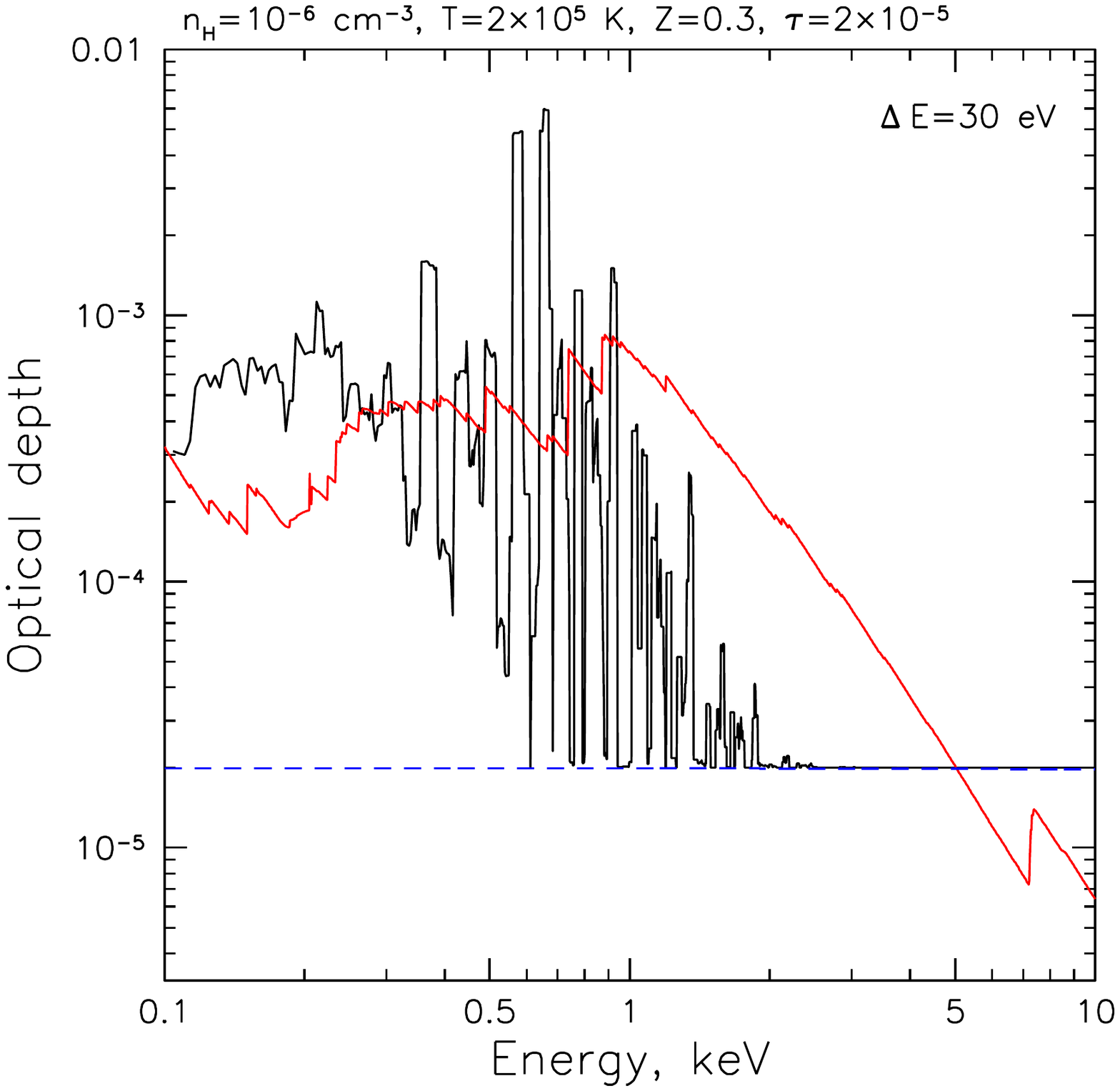}
\includegraphics[bb=50 180 600 700,width=0.31\textwidth]{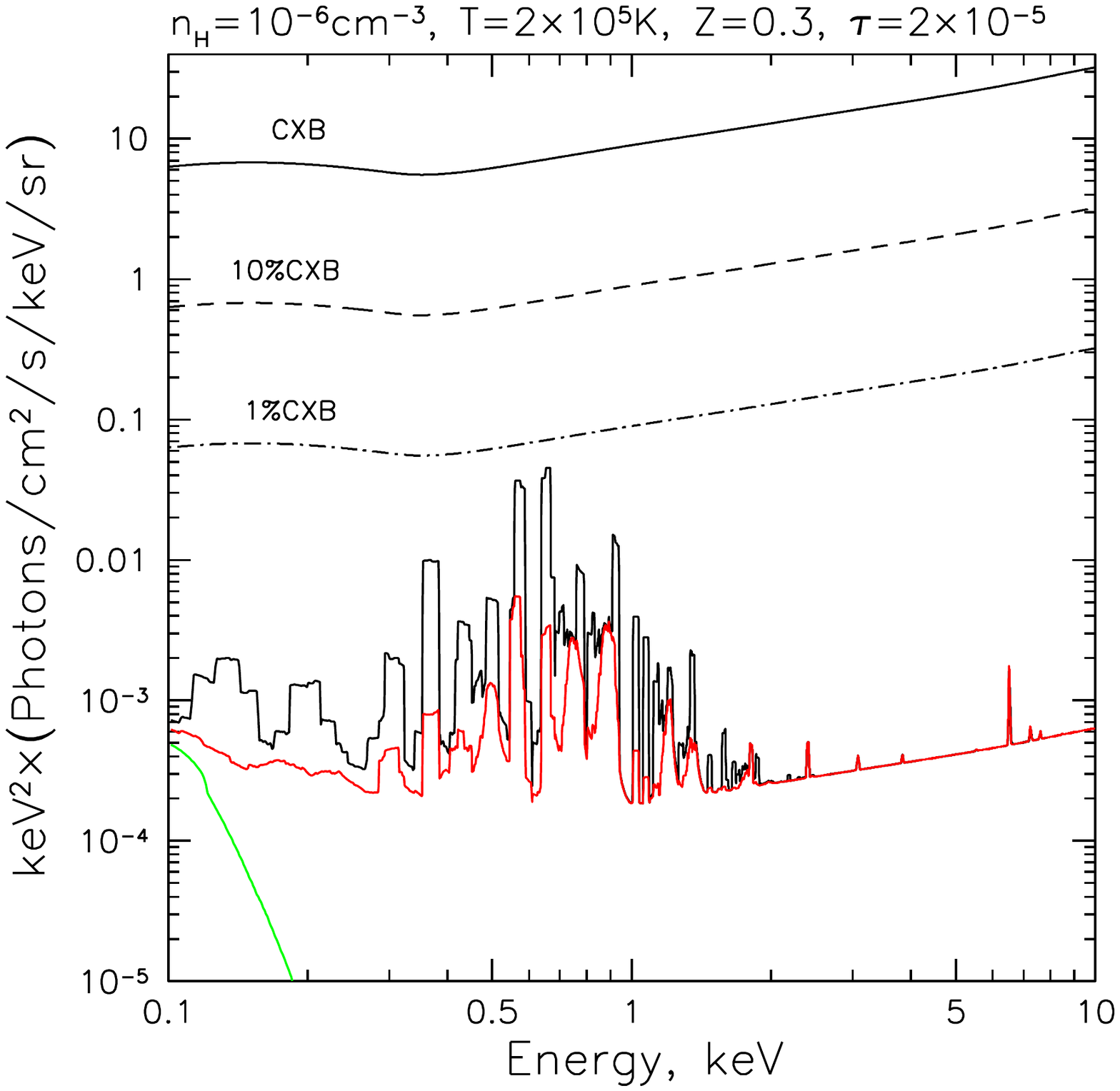}
\includegraphics[bb=50 180 600 700,width=0.31\textwidth]{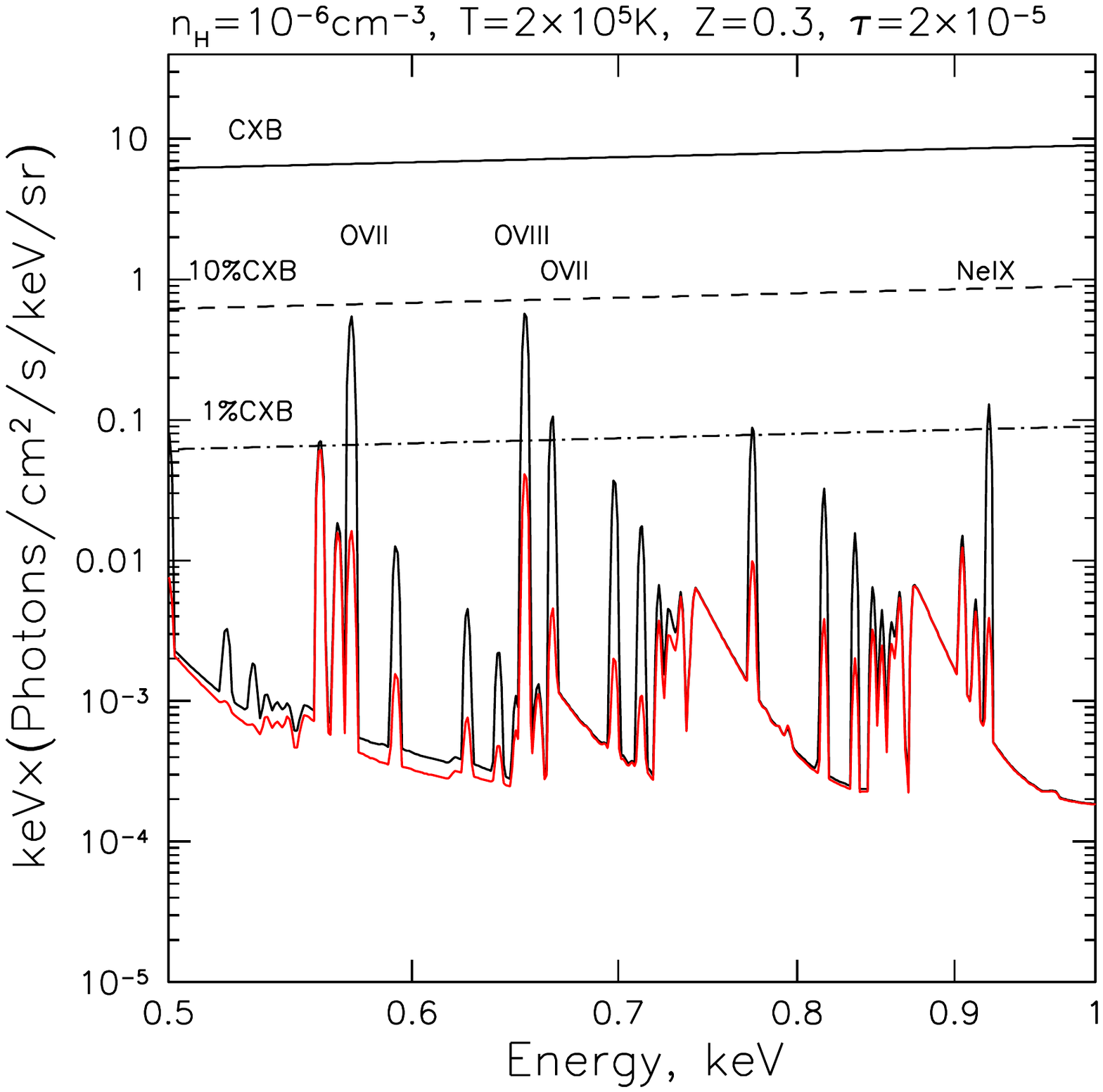}
\caption{\textbf{Left.} Scattering (black) and photo-absorption (red) optical depths of a plane parallel slab with $n_{H}=1\times10^{-6}$ cm$^{-3}$, T$=2\times 10^5$ K, Z=0.3, given that Thomson optical depth (blue dashed line) is $\tau_{e}=2\times 10^{-5}$. The resonant scattering optical depth was smoothed with a $ \Delta E=30$ eV-wide boxcar window. \textbf{Middle.} Broad band energy spectrum of the emission from such a slab with (black) and without (red) contribution from the resonantly scattered CXB included (also smoothed with a 30 eV-wide window). Emission predicted in the purely collisional case is shown with green. Intensity of the CXB, along with its 10\% and 1\% fractions are shown with solid, dashed and dash-dotted curves, as indicated next to each curve. \textbf{Right.} The 0.5-1 keV portion of the predicted photon spectrum for the same slab (with $ \Delta E=3$ eV boxcar smoothing applied). }
\label{fig:odsp_sh}
\end{figure*}

\begin{figure*}
\centering
\includegraphics[bb=50 180 600 700,width=0.31\textwidth]{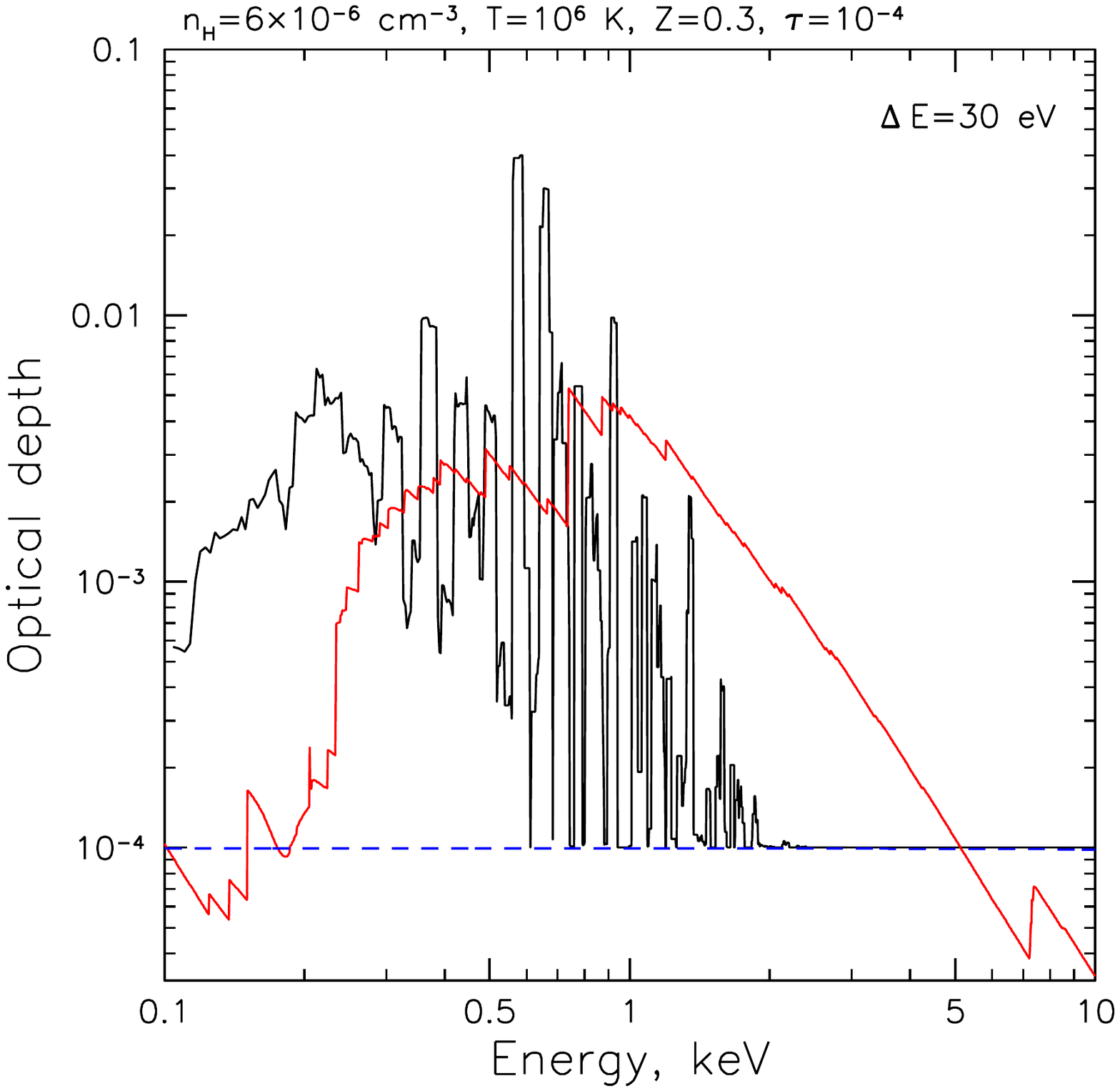}
\includegraphics[bb=50 180 600 700,width=0.31\textwidth]{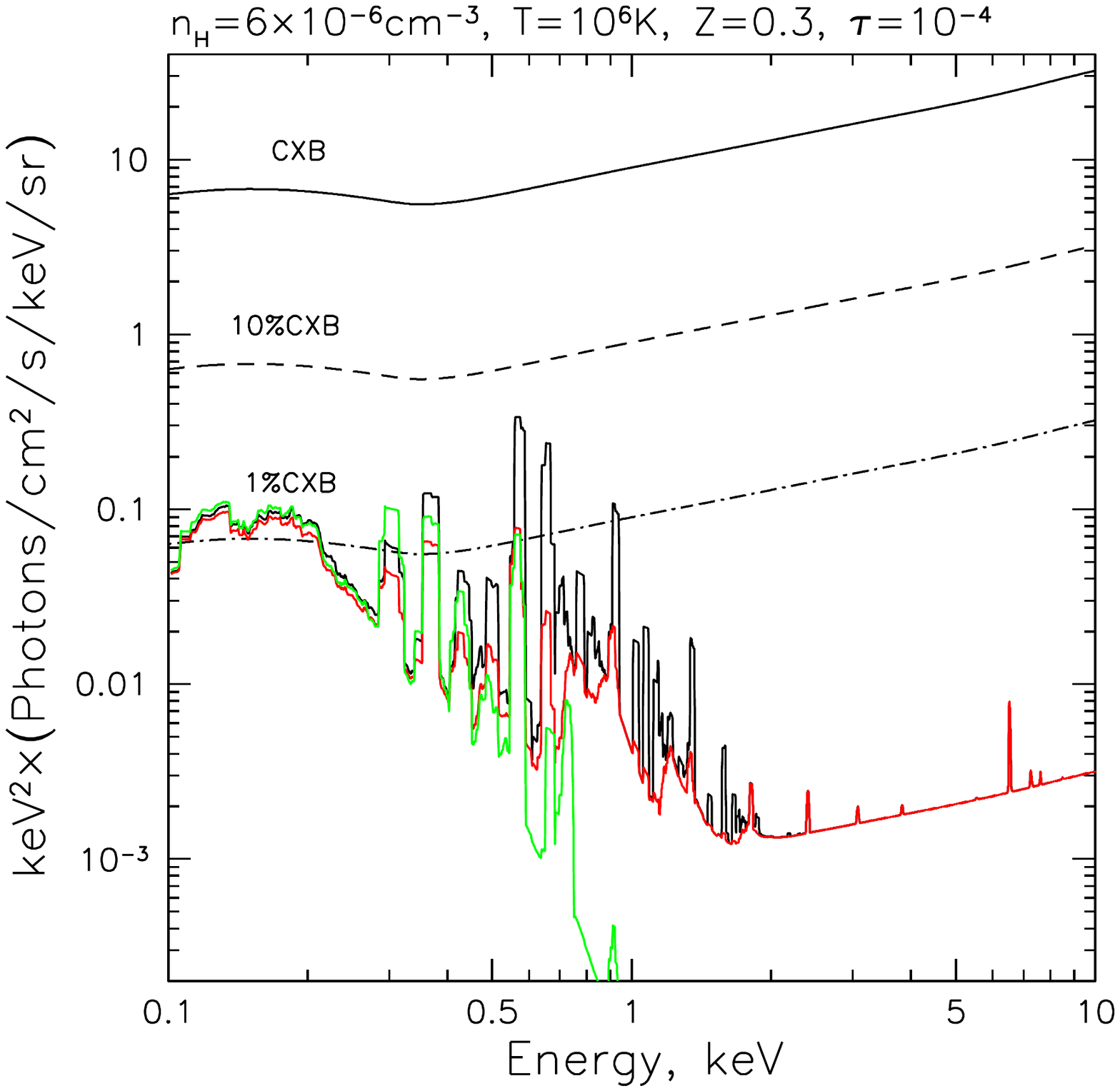}
\includegraphics[bb=50 180 600 700,width=0.31\textwidth]{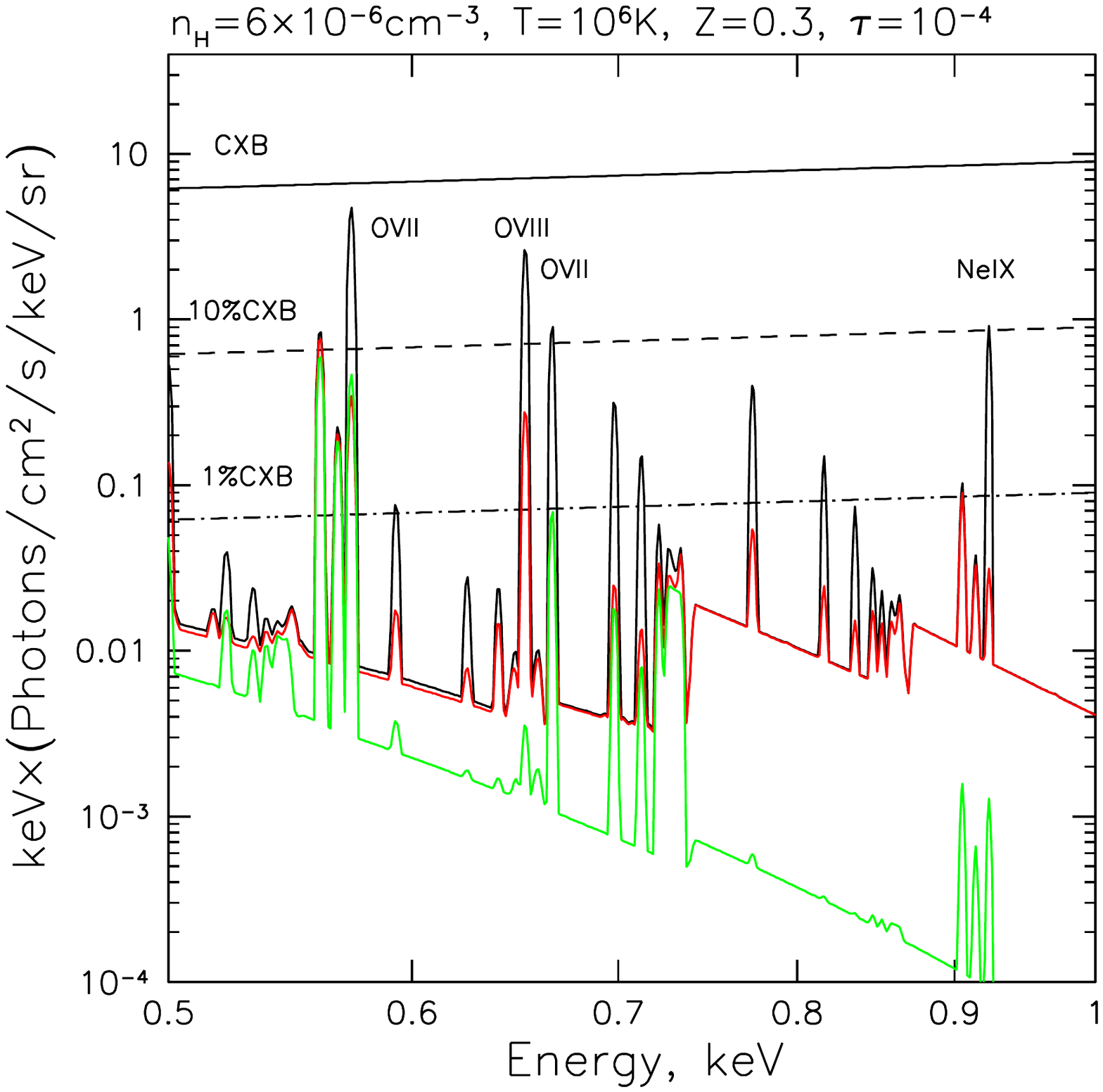}
\caption{The same as Fig.\ref{fig:odsp_sh} but for a denser and hotter slab with $n_{H}=6\times10^{-6}$ cm$^{-3}$, T$=10^6$ K.}
\label{fig:odsp_fi}
\end{figure*}

The resulting optical depths due to resonant scattering and photo-absorption for  sheet-like and filament-like structures are shown on left panels in Fig.\ref{fig:odsp_sh} and Fig.\ref{fig:odsp_fi}, respectively. The corresponding broad-band (0.1-10 keV) and soft X-ray (0.5-1 keV) spectra of their emission with (black) and without (red) account for resonant scattering are shown in the middle and the right panels in the same figures. In both cases the predicted emission is well above the purely collisional prediction, reflecting the great impact of CXB on the ionization state and X-ray emissivity of the WHIM. 

Evidently, the predicted continuum emission is 3-4 orders of magnitude smaller then CXB, and above 2 keV it is actually dominated by Thomson scattered CXB in both cases.  The line emission, however, appears very prominent for brightest lines between 0.5 and 1 keV, $\gtrsim 10$\% of CXB intensity for a filament and $\gtrsim 1$\% for a sheet parameters, if smoothed with a 3 eV-wide boxcar window (see right panels in Fig.\ref{fig:odsp_sh} and Fig.\ref{fig:odsp_fi}). {This reflects the fact that resonant optical depth in the line center turns out to be $\sim1.5$ for $ \tau_T\sim 10^{-4}$, while its intrinsic width is at $\sim$0.2 eV level (see Section \ref{ss:emission}).} Smoothing with a 30 eV-wide window lowers these values by a factor of 5-10, depending on how many adjacent bright lines get blended in such smoothing (compare the middle and right panels in Fig.\ref{fig:odsp_sh} and Fig.\ref{fig:odsp_fi}). As expected (see Section \ref{s:calculation}), the brightest lines are those of He- and H-like oxygen and neon (marked with OVIII, OVII and Ne IX in right panels of Figs. \ref{fig:odsp_sh} and \ref{fig:odsp_fi}). Below we quantify some characteristics of the predicted spectra in more detail and on the full grid of calculated models. 


\subsection{Intrinsic versus scattered emission}
\label{ss:ratio}

~~~~~Let us start with quantifying the boost in X-ray emissivity provided by resonant scattering compared to the intrinsic emission from WHIM. For an illustration, we will focus on the complex of lines of He-like oxygen around 0.57 keV, that appears to be one of the brightest for a wide range of parameter values. 

\begin{figure*}
\centering
\includegraphics[bb=50 150 620 700,width=0.31\textwidth]{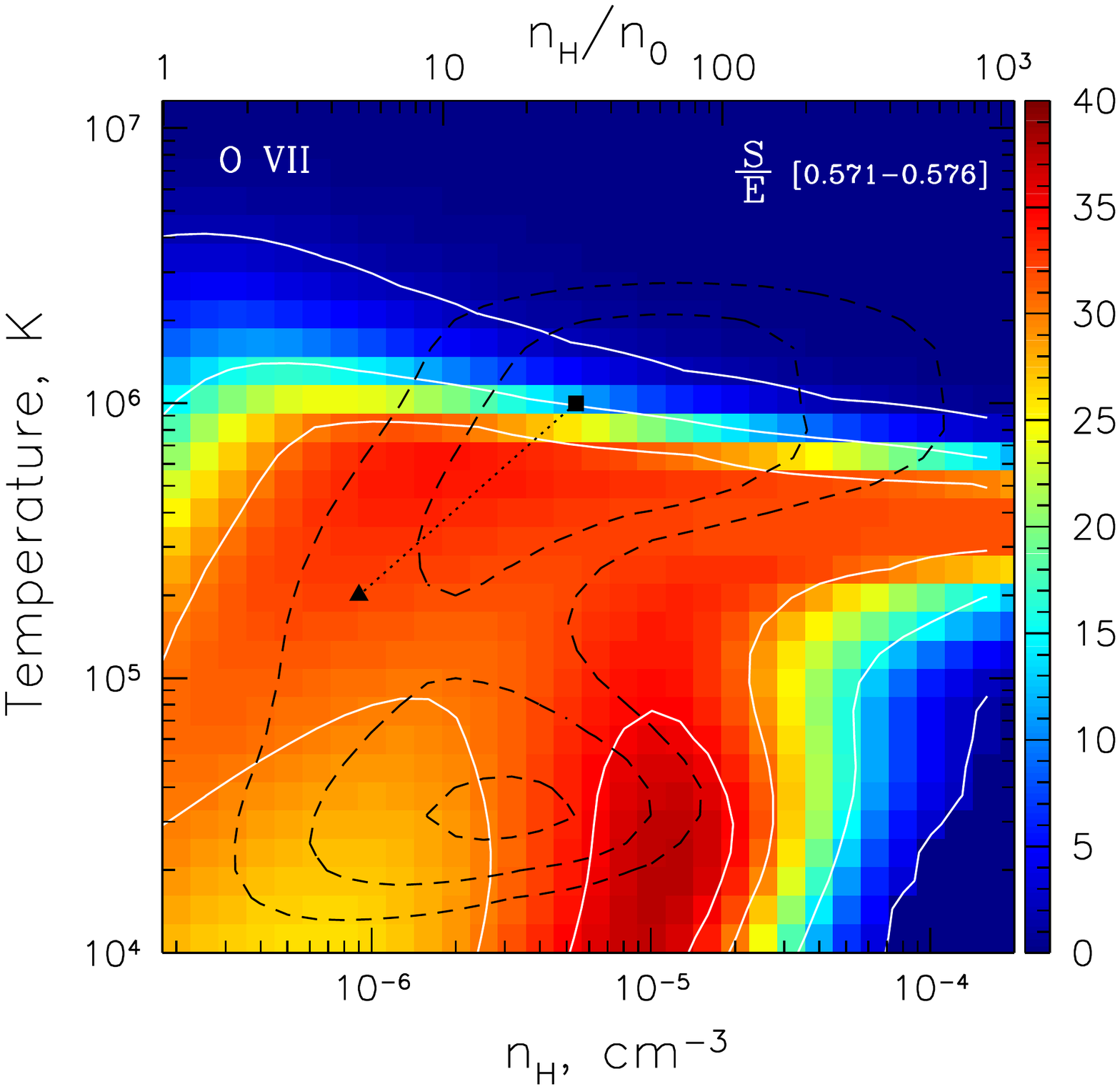}
\includegraphics[bb=50 150 620 700,width=0.31\textwidth]{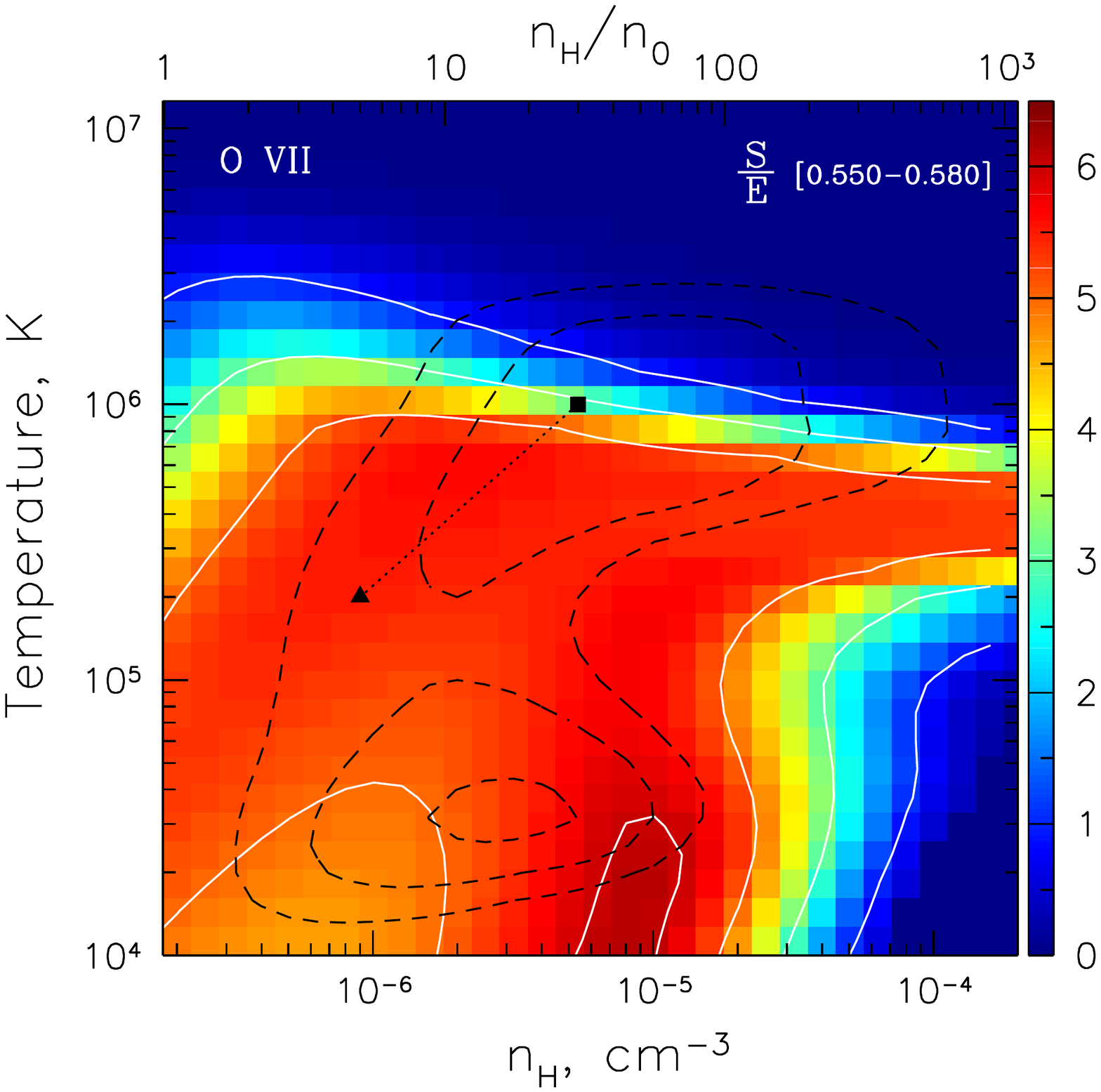}
\includegraphics[bb=50 150 600 700,width=0.31\textwidth]{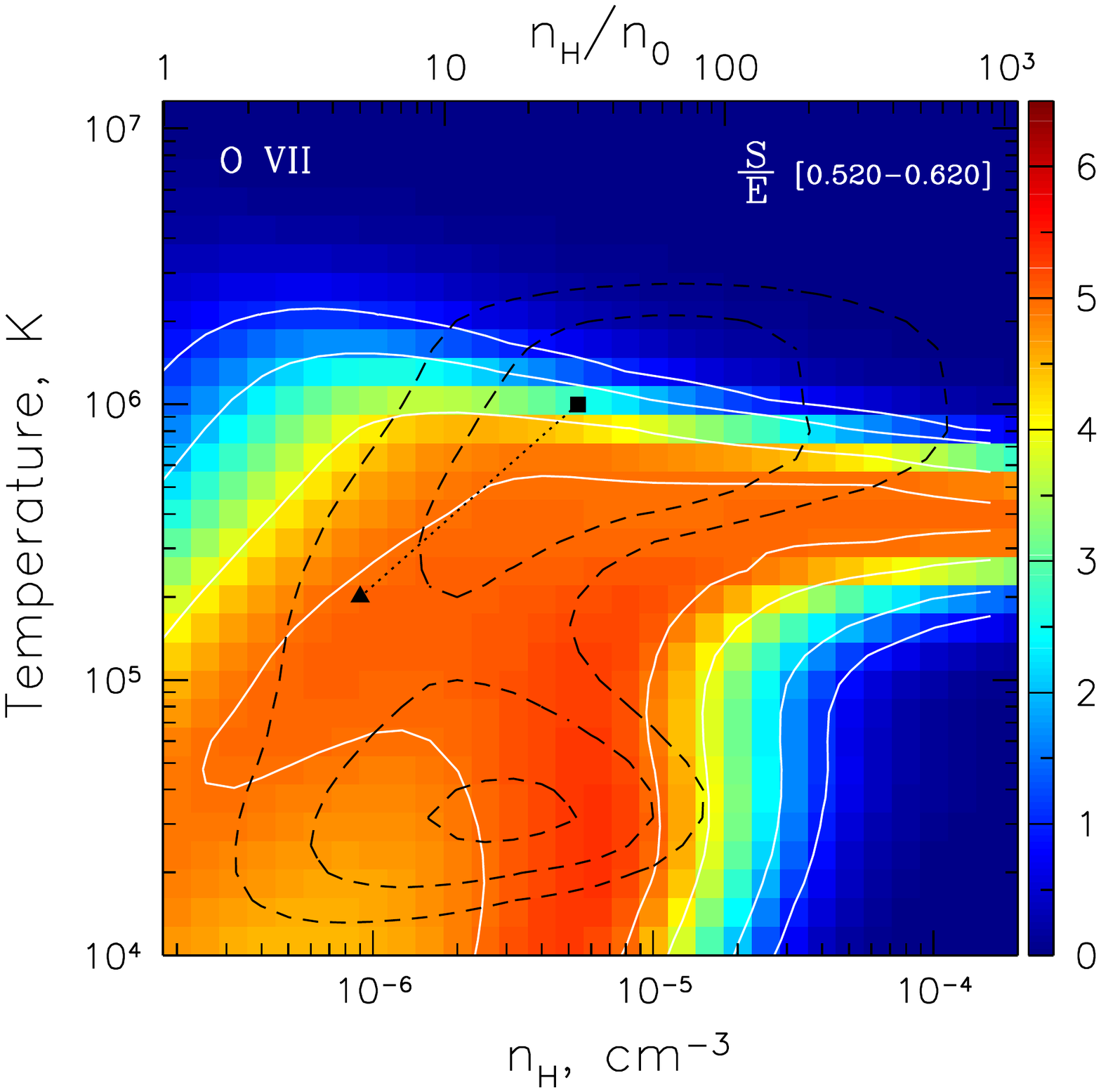}
\caption{Ratio of scattered to intrinsic X-ray emission from WHIM as a function of gas number density and temperature for three spectral bands: 0.571-0.576 keV (left panel, white solid contours at 2, 15, 30 and 35), 0.550-0.580 keV (middle panel, white solid contours at 1, 3, 5 and 6) and 0.520-0.620 keV (right panel, white solid contours at 1, 2, 4 and 5). {Note the different colorbar scale of the left panel. These bands correspond to 5-eV, 30-eV, and 100-eV-wide windows centred on the O VII resonant line, respectively.} The black dashed contours are the same as in Figure \ref{fig:dti_o_massmetion} and indicate O VII ionization fraction weighted with the mass and mean metallicity of the corresponding WHIM portion. The black solid triangle and square connected by a dotted line mark the parameters of sheet-like and filament-like structures illustrated in the text in more detail.}
\label{fig:scemrat}
\end{figure*}

First, we consider a very narrow (5 eV-wide) spectral band centred on the resonant line of this complex, viz. 0.571-0.576 keV band. In left panel of Fig.\ref{fig:scemrat} we show the ratio of scattered to intrinsic photon flux in this band, both of which are fully dominated by the resonant line emission, of course. In accordance with the estimates given in Section \ref{ss:emission} (see Eq. \ref{eq:scemtalin} and the discussion after it), the resulting ratio equals $\sim 30$ across a major fraction of the parameter space area relevant for WHIM (as demonstrated by the  white solid contours in the left panel of Figure \ref{fig:scemrat}). Clearly, for other strong resonant lines, like O VIII or Ne IX, the situation is pretty much similar {(see Appendix for the corresponding figures)}. Another important feature is a steep decline of this ratio at temperatures above $T\sim 10^6$ K for the whole range of considered densities due to decrease in ionization fraction of O VII. At lower temperatures, $T\sim$ few $10^4$ K, there is a slight local maximum on the ratio map at $n_{H}\sim 10^{-5}$ cm$^{-3}$ ($\delta \sim 50$), where OVI ionization fraction is high, and then there is a steep decline at higher densities. As a result, this ratio might be used as a diagnostic of temperature around $T\sim 10^6$ K across the whole range of densities, and as a density probe for temperatures between $10^4$ K and $10^5$ K and over-densities  $\delta \sim 100$.

Next, we consider a 30-eV window which contains the whole triplet of He-like oxygen, viz. 0.55-0.58 keV band. In addition to the resonant line, the triplet is composed of an intercombination and a forbidden line, for which the emissivity boost due to resonant scattering is much smaller or completely absent, respectively. Since for photoionized gas all components of the triplet have comparable amplitude in the intrinsic emission, the resulting ratio of scattered to intrinsic emission drops by a factor of $\sim5$ in this band, and it turns out to be between 5 and 6 for the bulk portion of WHIM (see the white solid contours in the middle panel of Figure \ref{fig:scemrat}). The overall morphology of this ratio's map is very similar to the narrower band ratio considered first, so it is capable of providing similar temperature and density diagnostic possibilities. Evidently, these diagnostics are less sensitive in this case, but they should be accessible with the spectral resolution at level $\sim$30-eV  already. {This is again true for the brightest O VIII and lines (see Figures \ref{fig:scemrat_oviii} and \ref{fig:scemrat_neix} in Appendix), though for O VIII the 30-eV-wide window encompasses also the K$_{\beta}$ line of O VII and doesn't contain bright forbidden lines, so the scattered-to-intrinsic ratio for the 30-eV-wide band is at almost the same level as for the 5-eV-band.  }

{The situation doesn't change strongly when 100-eV-wide windows are considered (see the right panel in Figure \ref{fig:scemrat} and Figures \ref{fig:scemrat_oviii} and \ref{fig:scemrat_neix} in Appendix). Such spectral windows roughly correspond to the typical spectral resolution of the CCD detectors like those on board \textit{XMM-Newton} and \textit{Chandra} X-ray observatories.}
   
\begin{figure}
\centering
\includegraphics[bb=50 150 620 700,width=0.48\textwidth]{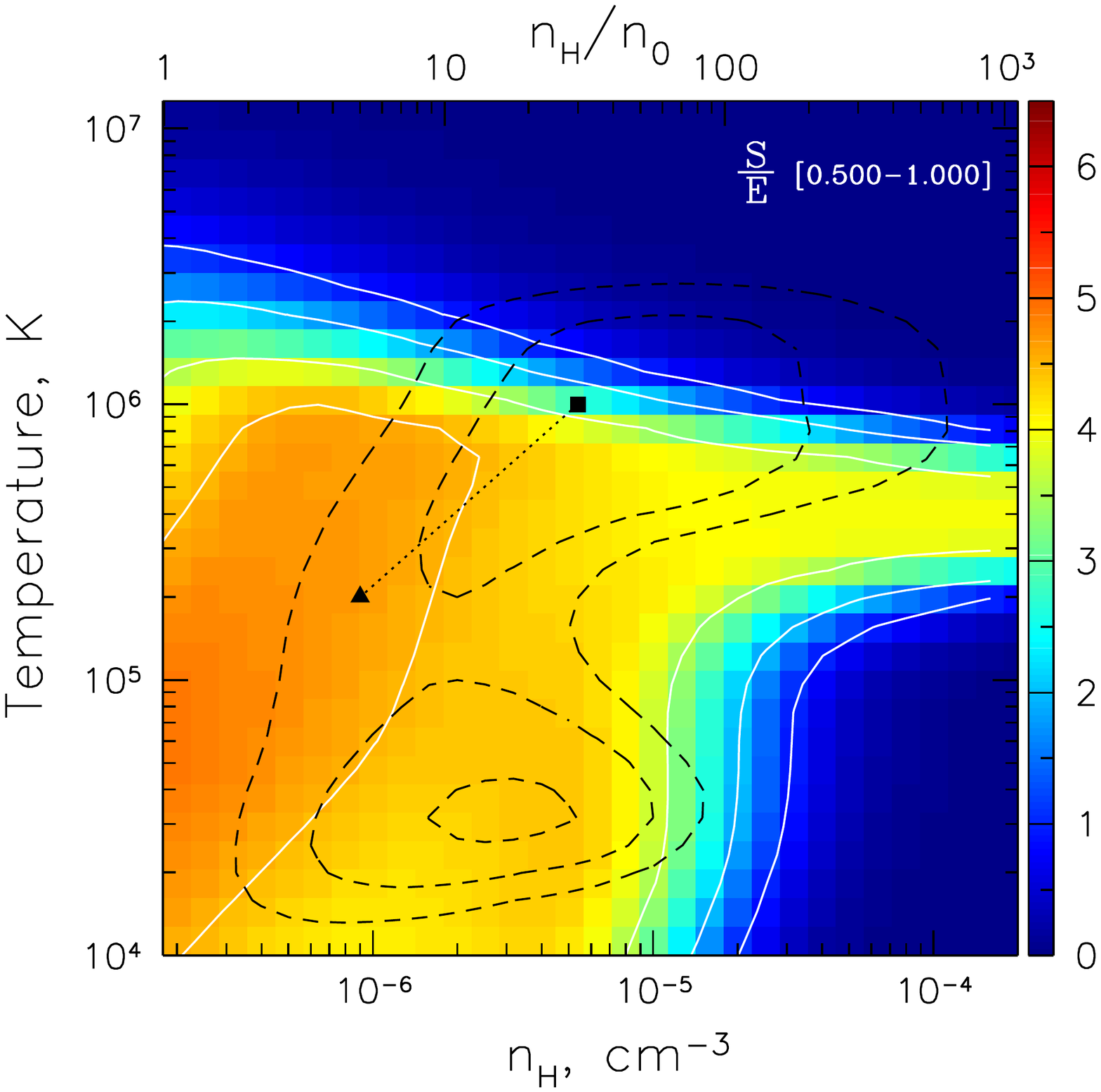}
\caption{Same as Figure \ref{fig:scemrat} but for the scattered and intrinsic emission integrated over the 0.5-1 keV spectral band (white solid contours at 1, 2, 3.5 and 4.5). The black dashed contours and marks are the same as in Figure \ref{fig:scemrat}.}
\label{fig:scemrat051}
\end{figure}

Finally, we calculate the scattered-to-intrinsic ratio for the whole 0.5-1 keV band, possessing majority of the brightest X-ray lines expected from WHIM. This ratio appears to behave similar to the ratios considered above {(see Fig. \ref{fig:scemrat051})}, and the bulk of WHIM has this ratio in the range 3.5-4.5. The overall morphology and locations of the most significant gradients is very similar to the ratios considered before, except for the absence of the peak at $T\sim$ few $10^4$ K and $n_{H}\sim 10^{-5}$ cm$^{-3}$, and the presence of a slight increase to the lowest densities. The resulting picture is pretty much consistent with the results presented in \citet{2001MNRAS.323...93C} as well (cf. Figure 4 there). The relatively small difference of this ratio compared to the ratio for the 30-eV-wide band reflects the fact that the resonant lines do contribute a very significant (if not dominant) fraction of the total X-ray emission in this band, as we demonstrate next.


\subsection{Line ratios}
\label{ss:lineratios}

\begin{figure*}
\centering
\includegraphics[bb=50 150 620 700,width=0.31\textwidth]{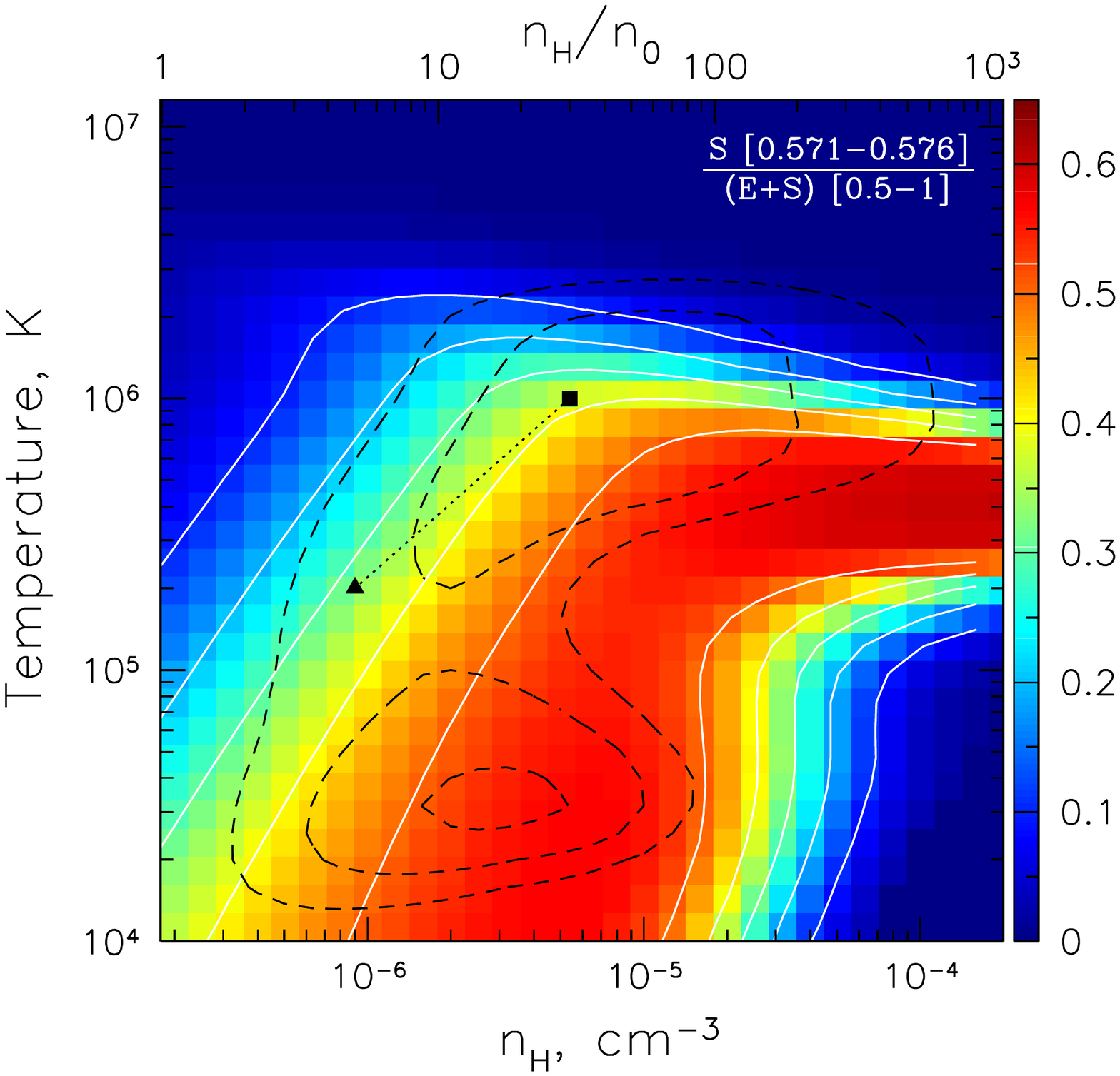}
\includegraphics[bb=50 150 620 700,width=0.31\textwidth]{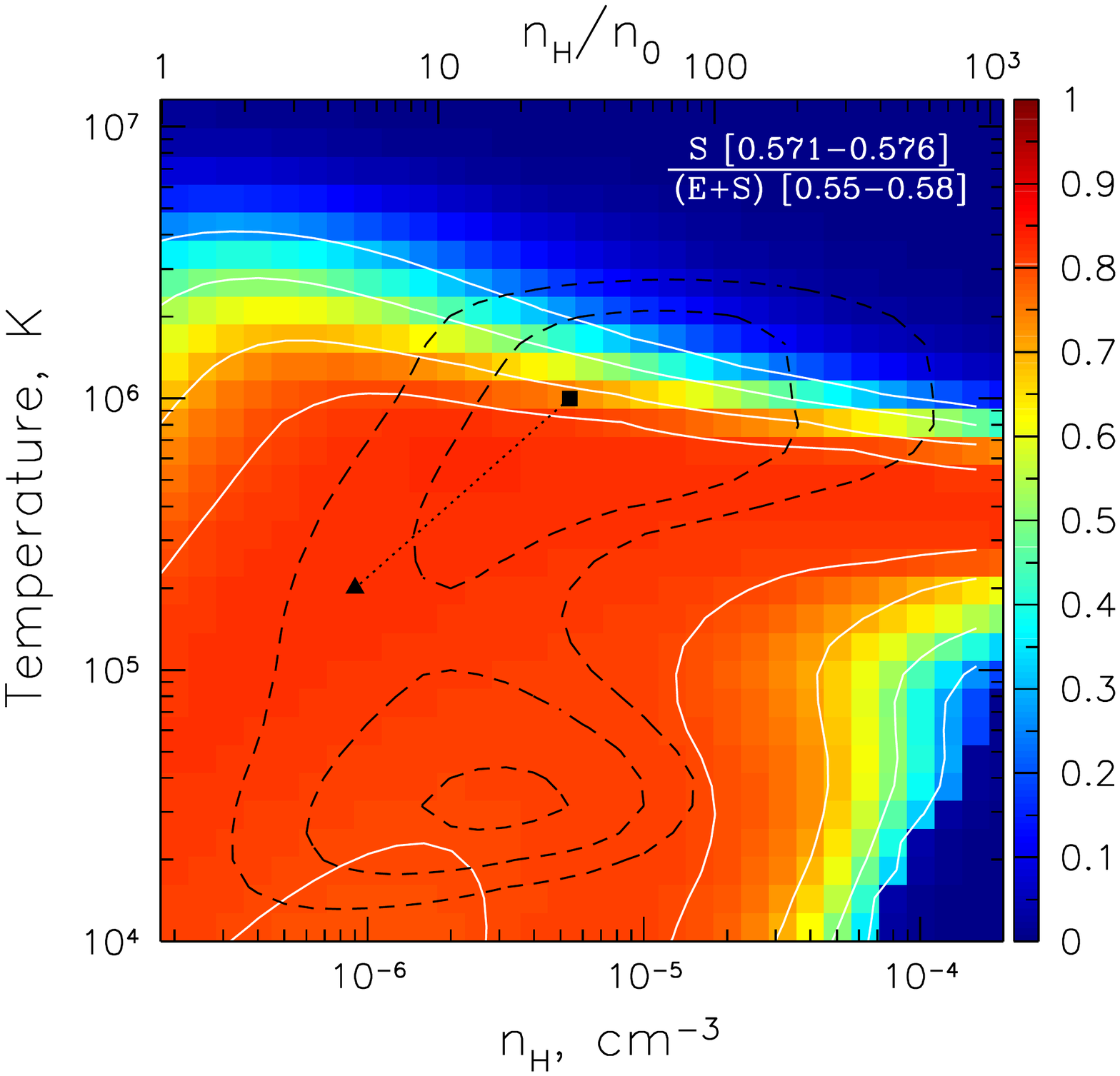}
\includegraphics[bb=50 150 620 700,width=0.31\textwidth]{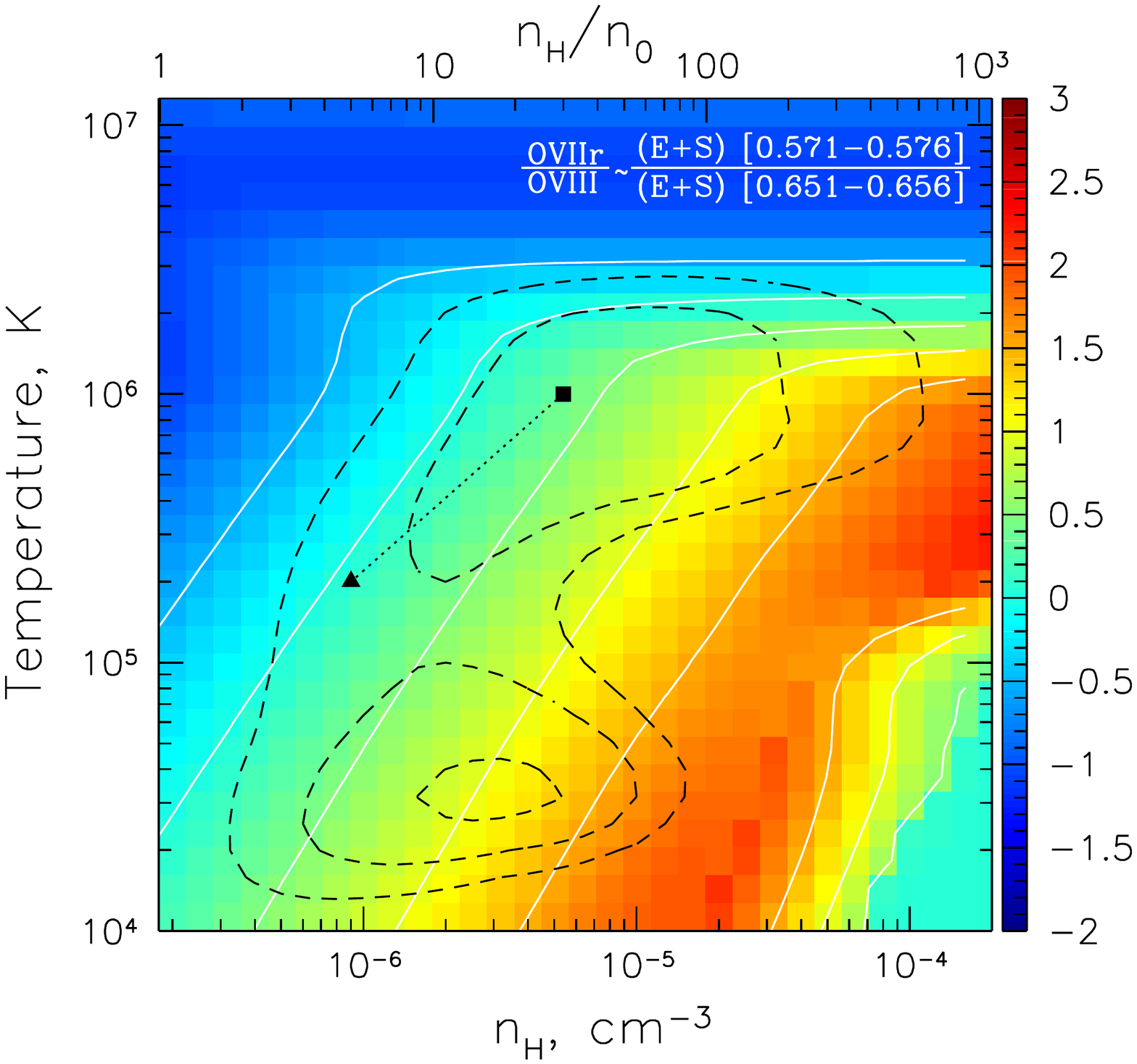}
\caption{\textbf{Left panel.} Ratio between scattered emission in 0.571-0.576 keV band (dominated by OVII resonant line) to the total (viz. scattered plus intrinsically emitted) radiation in 0.5-1 keV band. White solid contours are at 0.1, 0.2, 0.3, 0.4 and 0.5. \textbf{Middle panel.} The fraction of the total WHIM emissivity in the 0.55-0.58 keV band (which contains the full triplet of He-like oxygen) contributed by the  resonantly scattered component. White solid contours are at 0.3, 0.5, 0.7 and 0.8. \textbf{Right panel.}  \textit{Logarithm} of the ratio between total emissivity in resonant line of He-like oxygen at 0.574 keV to the total emissivity in the brightest resonant line of H-like oxygen at 0.653 keV. White solid contours are at -0.5, 0, 0.5, 1, 1.5. The black dashed contours, black solid triangle and black solid square connected by a dotted line in all panels are the same as in the previous plots.}
\label{fig:linrat}
\end{figure*}

In order to quantify the significance of the resonantly scattered emission line of He-like oxygen, we calculate the fraction of emissivity integrated over 0.5-1 keV that is contributed by this single line. Figure \ref{fig:linrat} (left panel) shows the map of the ratio between scattered emission in 0.571-0.576 keV band (dominated by OVII resonant line) to the total (viz. scattered plus intrinsically emitted) radiation in 0.5-1 keV band.  As can be seen from Fig. \ref{fig:linrat}, this single line contributes more than a half of total emissivity in the 0.5-1 keV band for an extensive region of the parameter space. However, this region doesn't cover the whole area relevant for WHIM, and there are actually significant gradients in this ratio in the directions of both lower and higher over-densities. The resulting picture in fully consistent with the results of calculations in \citet{2001MNRAS.323...93C}. Namely, both sheet-like and filament-like structures appear to lie between the 0.3 and 0.4 contour lines on this ratio's map. It is this ratio that should be relatively easiest to measure, and we clearly see that it has high diagnostic power for the gas parameters relevant to WHIM.

Next, we consider the fraction of the total WHIM emissivity in the 0.55-0.58 keV band, which contains the full triplet of He-like oxygen, that is contributed by the resonantly scattered component. This fraction turns out to be almost constant at level $\approx 0.8$ across almost the whole WHIM-related region of parameter space (see Fig.\ref{fig:linrat}, middle panel). That means, ratio of the sum of the forbidden and intercombination components to the resonant one should universally be around 0.1-0.2 for the He-like triplet of oxygen almost independently on the physical conditions in WHIM. As described in Section \ref{s:calculation}, this fact follows from the relation between resonant scattering and photo-ionization cross-sections and smoothness of the incident continuum.

Finally, let us consider the ratio between total emissivity in resonant line of He-like oxygen at 0.574 keV to the total emissivity in the brightest resonant line of H-like oxygen at 0.653 keV. \textit{Logarithm} of the resulting ration is shown in the right panel of Fig.\ref{fig:linrat}. Of course, this ratio varies much more significantly than others simply as a result of variations in the OVII and OVIII ionization fractions. For the bulk of the WHIM-related region, this ratio is between 1 and 10, but it falls below unity at lower densities and at $T\gtrsim 10^6$ K, of course. Interestingly, for both sheet-like and filament-like parameters, this ratio appears to be between 1 and 3, so both lines should be of comparable amplitude. This can also be seen directly in the spectral plots Fig. \ref{fig:odsp_sh} and Fig. \ref{fig:odsp_fi} (right panels). Combining this with the left panel of Fig.\ref{fig:linrat}, one can see that these two lines should contribute almost all emissivity in the 0.5-1 keV band for such parameters. That also means, that the corresponding weighted mean of photon energies in this band should be $\approx(0.574+0.653)/2=0.614$ keV, that should clearly affect redshift estimation from such emission (if the lines are not spectrally resolved). The corresponding redshift space distortion due to variation in relative contribution of O VII and OVIII line might amount to $\sim$few$\times$0.01. We discuss this and other observationally-important consequences of our study in the next Section.

\section{Discussion}
\label{s:discussion}

In the previous two sections we calculated and described basic properties of the total X-ray emission from WHIM, that includes contributions from intrinsically emitted radiation and resonantly scattered cosmic X-ray background. In this Section, we discuss observational implications of these predictions.

As we outlined in Section \ref{s:introduction}, the observational technique aimed at detection of WHIM in resonantly scattered emission relies on the possibility to resolve a significant fraction $ f $ of CXB into point sources {and treat separately absorption features in spectra of the resolved sources and the emission features in residual diffuse signal.} It is useful to notice that angular size of a structure with size $L=10~h^{-1}$ Mpc equals 2.15, 1.2 and 0.9 deg (i.e. 130, 72 and 54 arcmin) at z=0.1, 0.2 and 0.3, respectively (for the reference cosmological model cited in Section \ref{s:introduction}). That means such structures are larger or comparable to the field-of-view (FoV) diameter of the currently operating (\textit{Chandra} and \textit{XMM-Newton}) and forthcoming (\textit{Spectrum-RG}/eROSITA\footnote{\href{http://erosita.mpe.mpg.de/}{http://erosita.mpe.mpg.de/}}) X-ray missions. Hence, one might assume that intrinsic and scattered emission from such structures would fill the whole FoV of such telescopes. 

Sensitivity to point sources at levels $10^{-14}$ and $10^{-15}$ erg s$^{-1}$ cm$^{-2}$ in the 0.5-2 keV band ensures resolving $\sim$60\% and $\sim$80\% of the CXB, respectively \citep{2003ApJ...588..696M}. Further increase is hard to achieve because of flatness of the resolved fraction curve at low fluxes \citep{2003ApJ...588..696M}. Since the amplitude of the scattered emission is linearly proportional to $f$, the difference between $f=0.8$ and $f=1$ is, however, only 20\%. Of course, increasing $f$ is also needed to suppress the Poisson noise produced by the residual unresolved CXB, which is proportional to $\sqrt{1-f}$. However, the contribution of this noise should be subdominant with respect to the contamination due to foreground soft X-ray emission of the hot gas in our own Galaxy \citep{2002A&A...389...93L}. This emission is diffuse in its nature, so it can not be efficiently filtered out from the aperture. One has to rely on some model of its spatial and spectral distribution, and allow for possible statistical (due to Poisson noise) and systematic (due to variation in line-of-sight absorption or presence of brighter clumps) fluctuations of it.     

Spectral shape of this emission is commonly modelled as a combination of two thermal optically thin components with temperatures $\sim0.2$ keV and $\sim0.074$ keV \mbox{\citep{2002A&A...389...93L}}. {While recent studies \citep[e.g. ][]{2017ApJ...834...33L} suggest slightly different spectral approximation of the softer component in this model (that is commonly attributed to the Local Hot Bubble emission), for our purpose it is more convenient to use the \citet{2002A&A...389...93L} results that describe the total Galactic X-ray background. It is worth mentioning, however, that the Local Hot Bubble emission can vary by a factor of several from place to place, and this variation must be properly included in the uncertainty budget of the Galactic X-ray background modelling.}

\begin{figure*}
\centering
\includegraphics[bb=50 180 600 700,width=0.32\textwidth]{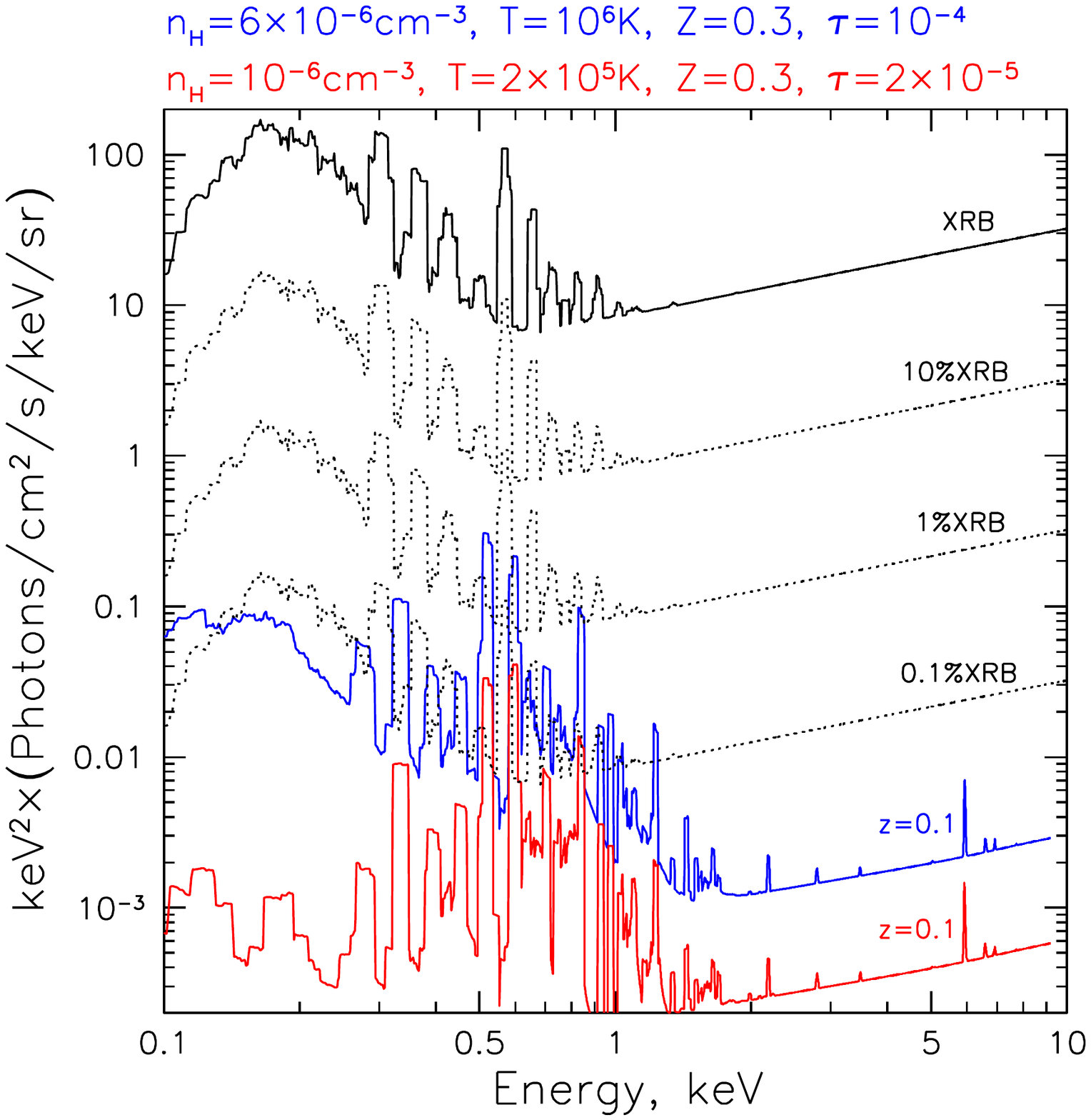}
\includegraphics[bb=50 180 600 700,width=0.32\textwidth]{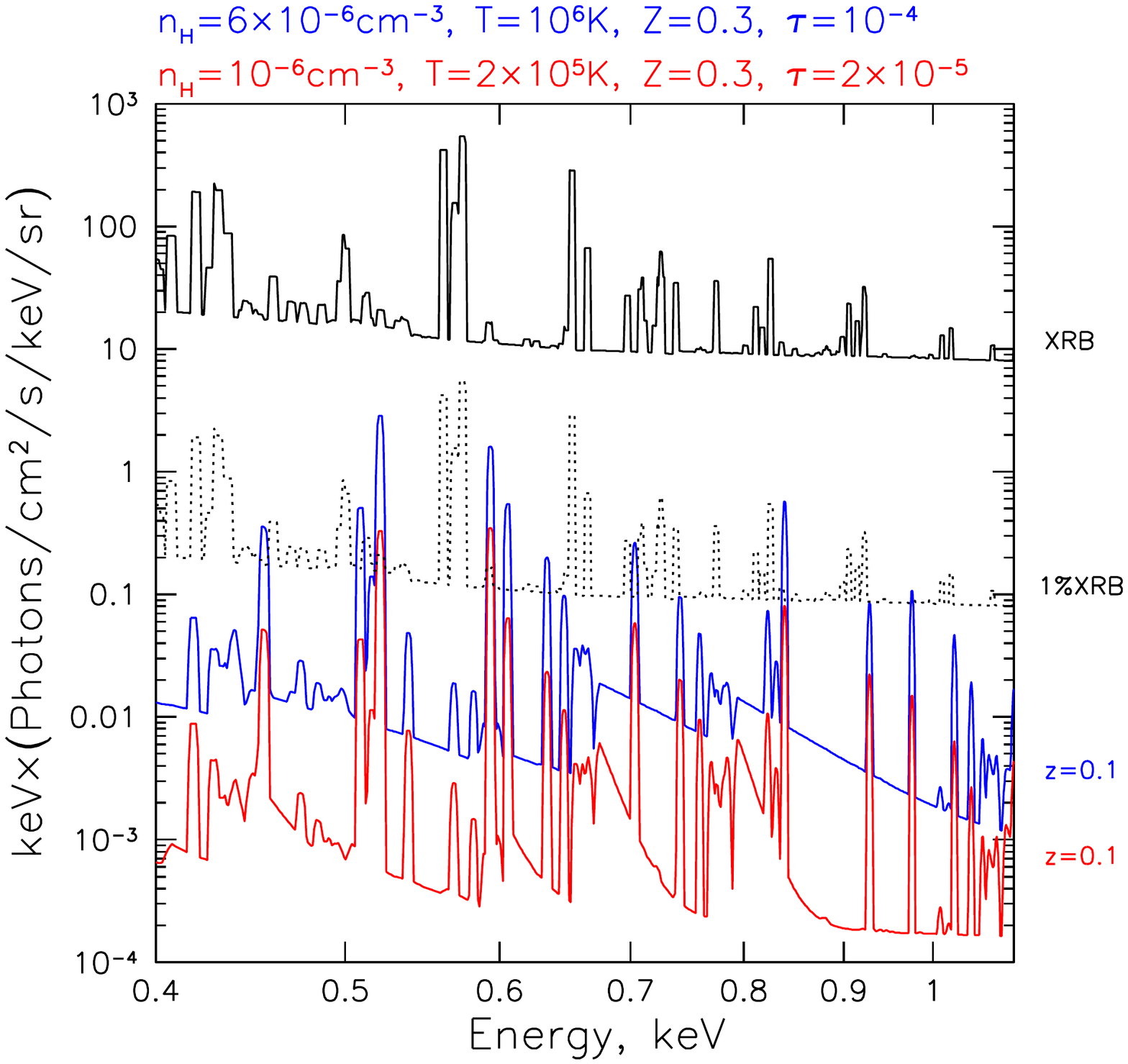}
\includegraphics[bb=50 180 600 700,width=0.32\textwidth]{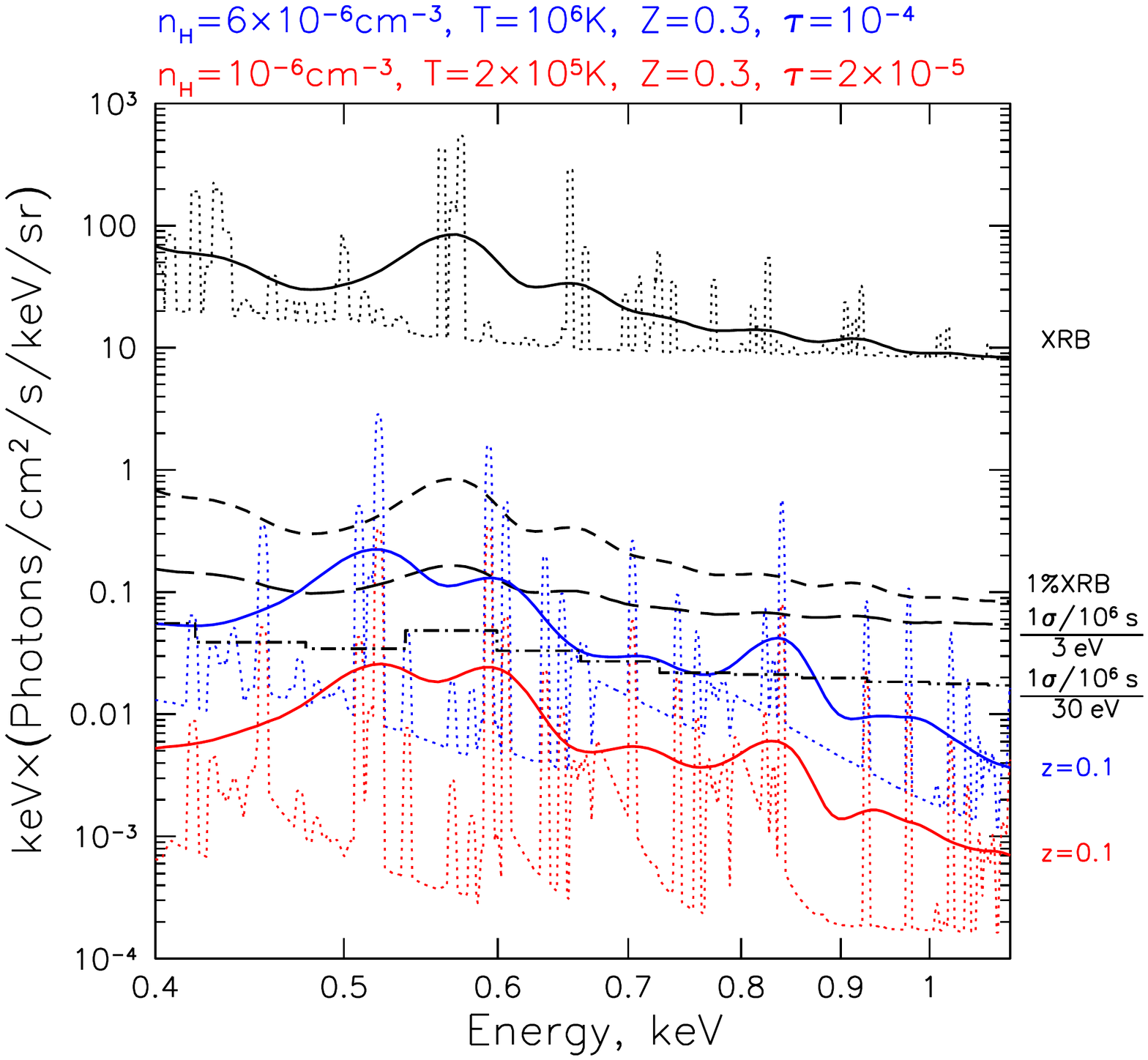}
\caption{\textbf{Left.} Broadband energy spectrum of the total (intrinsic plus scattered) X-ray emission from a filament-like (blue) and a sheet-like (red) structures located at redshift $z=0.1$ in comparison with the full X-ray background radiation (solid black, cosmic X-ray background plus Galactic X-ray emission). Parameters of these structures are indicated on top of the figure. All spectra were smoothed with a 30-eV-wide boxcar window. \textbf{Middle.} Photon spectrum of total emission from the same structures in 0.4-1.2 keV band (smoothed with a 5-eV-wide boxcar window), where the most prominent emission lines in WHIM emission (O VII, O VIII and Ne IX) are located. \textbf{Right.} The same as middle panel (dotted curves) but with addition of spectra convolved with the response function of \textit{SRG}/eROSITA detectors. Long-dashed (3-eV-wide bin) and dash-dotted (30-eV-wide bin) lines show the 1$\sigma$ noise level per bin, as predicted for a $10^6$s-long observation of structures fully filling field-of-view of the telescope.}
\label{fig:spxrbfish}
\end{figure*}

{Both components of the Galactic X-ray background} posses strong emission lines, including the same lines of O VII and O VIII that are also the strongest ones in the predicted emission from WHIM (see Fig. \ref{fig:spxrbfish}). This significantly hampers detection of filaments at $z\sim0$.  For structures at  $z\sim0.1$, however, the Galactic O VII emission line falls in between O VII and O VIII lines from WHIM, so the overlap and corresponding contamination are significantly reduced even for spectra with $\sim50$-eV resolution (see Fig. \ref{fig:spxrbfish}). 

Let us estimate the level of noise one might expect to achieve with an X-ray instrument having FoV area $A=1$ deg$^{2} A_1$, effective area at 0.5 keV $S(0.5 \mathrm{keV})=1000$ cm$^{2} S_{1000}$ and spectral resolution at level $\sigma\sim50$-eV. For the CXB radiation field described in Section \ref{ss:radiation}, one has $ I_{\mathrm{CXB}}(E)\sim 10^{-2}$ ph $^{-1}$ s$^{-1}$ cm$^{-2}$ keV$^{-1}$ deg$^{-2}$ at E=0.5 keV. Hence, it will produce $\sim 1\times A_1 S_{1000}\Delta E_{100}$ counts per second in a 100-eV-wide band pass $\Delta E$ around 0.5 keV across the whole field of view (contribution of the Galactic foreground emission is discussed in the next paragraph). If the fraction of CXB that is resolved into points sources and excluded from the aperture equals $f=0.8$, the remaining unresolved part would give $\sim 2\times10^5\, \left(\frac{1-f}{0.2}\right) A_1 S_{1000}\Delta E_{100}\Delta t_{Ms} $ counts detected in such a band during an exposure time of $\Delta T=10^6 \times\Delta t_{Ms}$ seconds. The resulting Poisson noise produced by this unresolved component is $\sim 500 \sqrt{\left(\frac{1-f}{0.2}\right) A_1 S_{1000}\Delta E_{100}\Delta t_{Ms}}$ counts, that is $\sim 4\times 10^{-4}$ of the full CXB. 

The Poisson noise due to foreground Galactic emission is a factor of $\sim$few$/\sqrt{1-f}\sim 5$ higher, with significant spectral variation due to contribution of the emission lines. Thus, one might expect the statistical noise to be at level of $2\times 10^{-3}$ CXB in a 100-eV wide band around 0.5 keV for the considered configuration. For a 30-eV wide passband, it should be a factor of 2 higher, i.e. at level of  $4\times 10^{-3}$ CXB. As we have shown in Section \ref{s:results}, the predicted X-ray emission from WHIM might exceed 1\% of the CXB level for a filament-like structure even after smoothing with a 30-eV-window, hence such structures (with $\tau_{T}\sim 10^{-4}$) would be detected at $\sim 10\sigma$ significance. For sheet-like structures, only marginal ($\sim 1\sigma$) detection would be possible, however.

The parameters of an X-ray instrument we used above match very well parameters of the eROSITA telescope on board \textit{Spectrum-RG} mission \citep{2016SPIE.9905E..1KP}. To illustrate this point, we simulated a $10^6$s-long observation for a filament-like and a sheet-like structures at $z=0.1$ filling the whole FoV of eROSITA (1 deg in diameter). For doing that, we used FoV-averaged response functions and the X-ray background model by \citet{2002A&A...389...93L}.  The resulting spectra and corresponding noise levels for 3-eV-wide and 30-eV wide binning are shown in Fig. \ref{fig:spxrbfish} (right panel). Filament-like structures might be detected even with 3-eV-wide binning but only around the O VII line, while for 30-eV-wide binning significant detection might achieved for around O VII, O VIII and Ne IX lines. For a sheet-like structure, only marginal detection might be feasible for the oxygen's pair of lines even with 30-eV-wide binning. 

{Of course, in reality the situation is more complicated since the modestly over-dense structures can be spatially
correlated with much denser and hotter structures, like galaxy clusters and groups. Intrinsic emission
from the latter structures is characterized by approximately the same redshift, while being much brighter,
of course, so that it can easily contaminate the WHIM signal. The structures which are most relevant for the
current study are, however, those of the over-density in the range  $\delta \sim 5-50$, which can in principle be
found far enough from the bright virialized objects to mitigate the contamination problem. Clearly, the corresponding observing strategy should imply deep and extensive coverage of relatively empty (in terms of the bright galaxy clusters) areas, which will be accessible with the future wide-field X-ray observatories.}

Detection with finer spectral binning is of great importance due to several reasons. First, possibility to identify individual lines allows one to measure the redshift of the WHIM structure, since position of the line centroid can in principle be measured with high accuracy even with a low resolution spectrometer. The redshift information is very useful for potential cross-correlation of the diffuse X-ray signal with high resolution spectroscopy of the resolved sources (available with grating spectrometers) or other tracers of the large scale structure. Also, redshift information allows some physical parameters of the WHIM gas to be derived from the measured size, flux and, perhaps, line ratios of the detected emission (see Section \ref{s:results}). Second,  finer spectral binning is needed to distinguish WHIM structures located at different redshifts along the line-of-sight. Evidently, one cannot distinguish structures separated in velocity space by less then the spectral resolution of the instrument, which amounts to $\delta z\sim 0.1$ for \textit{XMM-Newton}, \textit{Chandra} or \textit{SRG}/eROSITA detectors. Hence, they in principle allow separating structures in the Local Universe from more more distant objects at redshifts $\gtrsim 0.2$, given significantly high signal-to-noise ratio for each individual structure.  

Future X-ray mission equipped with X-ray calorimeters (e.g. \textit{ATHENA}\footnote{\href{http://www.the-athena-x-ray-observatory.eu/}{http://www.the-athena-x-ray-observatory.eu/}}) will provide both higher sensitivity for individual structures and better possibility to separate them in the redshift space, so that measurement of the WHIM correlation length might become feasible even from small patches of the sky. This is of course very similar to the studies of X-ray absorption forest in the spectrum of the resolved background point sources (a direct analogue of Lyman-$\alpha$ absorption forest in spectra of high-redshift quasars), and these two techniques can be combined to infer relative contribution of the intrinsic (not scattered) emission. The ratio of scattered-to-intrinsic emission is a direct probe of gas properties, as we have shown in Section \ref{s:results}. The relative line intensities might be the used to further constrain some parameters, and check validity of the ionization equilibrium assumption \citep[e.g.][]{2006PASJ...58..641Y}.

Although the actual intensity of the WHIM emission from individual structures might, of course, significantly vary because of variations in integrated column density and mean metallicity, our predictions demonstrate detection feasibility on average. This has particular relevance for the \textit{SRG}/eROSITA mission, which primary goal is performing an all-sky survey with average exposure $t_{exp}\approx 2$ ks and point sources sensitivity $\approx 10^{-14}$ erg s$^{-1}$ cm$^{-2}$ in the 0.5-2 keV band \citep{2012arXiv1209.3114M}. This will allow resolving $\sim 60$\% of CXB over the whole extragalactic sky, i.e. over area of $A\sim 20 000$ deg$^2$. The corresponding `grasp' of this survey is $A\times t_{exp}\sim 4\times 10^7$ deg$^2\times$s, i.e. 40 times more than of single $10^6$s-long pointing. That means, if the fraction of the sky covered by WHIM structures with parameters similar to those considered here amounts to a few percent at $z\sim0.1$, stacking of the all-sky data would be equivalent to the single deep observation simulated above. In such a case, additional information might be provided by cross-correlation of this signal with various traces of the large-scale structure in the Local Universe (e.g. \textit{2MASS} galaxies). 

Another possibility  might be related to searching for a decrement in a stacked spectrum of the background points sources left to the positions of the O VII and O VIII lines, caused by resonant scattering and photo-ionization absorption edges integrated over the whole redshift range between the source and the observer.  In this case, Galactic foreground emission is less of an issue because of point-likeness of sources. which can also be grouped according to their redshift in order to probe the growth of the average WHIM scattering optical depth with redshift (this has some similarity to the situation observed in so-called Lyman-break galaxies). The all-sky stacking approach allows one to eliminate some systematics inherent in studies of individual structures. As a result, it would provide a direct measure of the total mass of the missing baryons residing in WHIM.       
 
%

\section{Conclusions}
\label{s:conclusions}
   
We revisited the calculation of the X-ray emission from WHIM including contribution of the resonantly scattered CXB emission. The latter should be taken into account given that significant fraction of CXB itself is resolved into point sources and excluded from the X-ray emission integrated over some aperture. We confirm the general conclusions of the previous study by \citet{2001MNRAS.323...93C}, and quantify some of the predictions in more detail.

The overall boost of emission in the most prominent resonant lines (O VII, O VIII and Ne IX) equals $\sim 30$, {and it is pretty much uniform across almost the whole region of the density-temperature diagram relevant for WHIM.} After averaging over broader spectral bands, it diminishes to $\sim 6$ for 30-eV wide smoothing and to $\sim 4$ when the full 0.5-1 keV band is considered. The ratio of the scattered to intrinsic emission, however, declines steeply at temperatures above $T\sim 10^6$ K for the whole range of considered densities and at over-densities $\delta \gtrsim 100$ for temperatures between $10^4$ K and $10^5$ K. The comparison between absorption in spectra of background point sources with the total emission from WHIM might be used as a diagnostic for gas properties in this regions of the parameter space.  

The resonantly scattered O VII line at 0.574 keV should dominate the total (scattered plus intrinsic) emission from the full triplet of He-like oxygen at 0.56-0.58 keV, contributing $\sim80$\% of the flux in this band for the bulk portion of the parameter space relevant for WHIM. Relative contribution of this single line to the total emission in 0.5-1 keV turns out to be $\sim 40$\%, with steep decline, however, at lower densities and higher temperatures, where O VIII takes over as oxygen's most abundant ionization specie. The two brightest lines of O VII and O VIII have comparable intensity for the major portion of WHIM, including the parameters of the reference sheet-like and filament-like structures, which we illustrated in more detail.
 
A significant detection of WHIM in emission might be achieved by an X-ray instrument with effective area $\sim 1000$ deg$^2$ at 0.5-1 keV after 10$^6$s-long exposure over an 1-deg$^2$ region of the sky, which is  fully covered by a filament-like structure (Thomson optical depth $\sim 10^{-4}$) at redshift $\sim0.1$. We demonstrate this by simulations of such an observation using the actual response functions of the \textit{SRG}/eROSITA telescope and the full X-ray background (cosmic X-ray background plus Galactic diffuse soft X-ray emission). 

Future X-ray missions provide great opportunities to study WHIM. This refers both to large area X-ray surveys and deep small area observations with X-ray calorimeters. For the former, the signal can be detected by a cross-correlation of stacked (absorption and emission) X-ray signal with certain tracers of overdensities in the large-scale structure, while for the latter detection (and potentially diagnostics) of the prominent individual filaments at $z\sim0.1$ is the primary goal. 
      
The calculated table model, suitable for use in numerical simulations and data analysis, is publicly available at \href{https://www.mpa-garching.mpg.de/~ildar/igm/}{https://www.mpa-garching.mpg.de/\~\,ildar/igm/}. {The extracted scattered-to-intrinsic emissivity ratios are also provided there in the form of FITS images. }
%
%
\section*{Acknowledgements}

We are grateful to Klaus Dolag for providing us with the baryonic mass and metallicity distributions extracted from the \textit{Magneticum} simulation. We acknowledge partial support by grant No. 14-22-00271 from the Russian Scientific Foundation. {We are grateful to the anonymous referee for the careful reading of the manuscript and helpful suggestions.}
\section*{Appendix}

{Here we present the scattered-to-intrinsic emissivity ratios for the 5-eV, 30-eV and 100-eV-wide spectral bands centred on the $K_{\alpha}$ lines of hydrogen-like oxygen O VIII and helium-like neon Ne IX. These lines are among the brightest lines in the 0.5-1 keV spectral band for typical WHIM conditions (see e.g. Figures \ref{fig:odsp_sh} and \ref{fig:odsp_fi}).}

\begin{figure*}
\centering
\includegraphics[bb=50 150 620 700,width=0.31\textwidth]{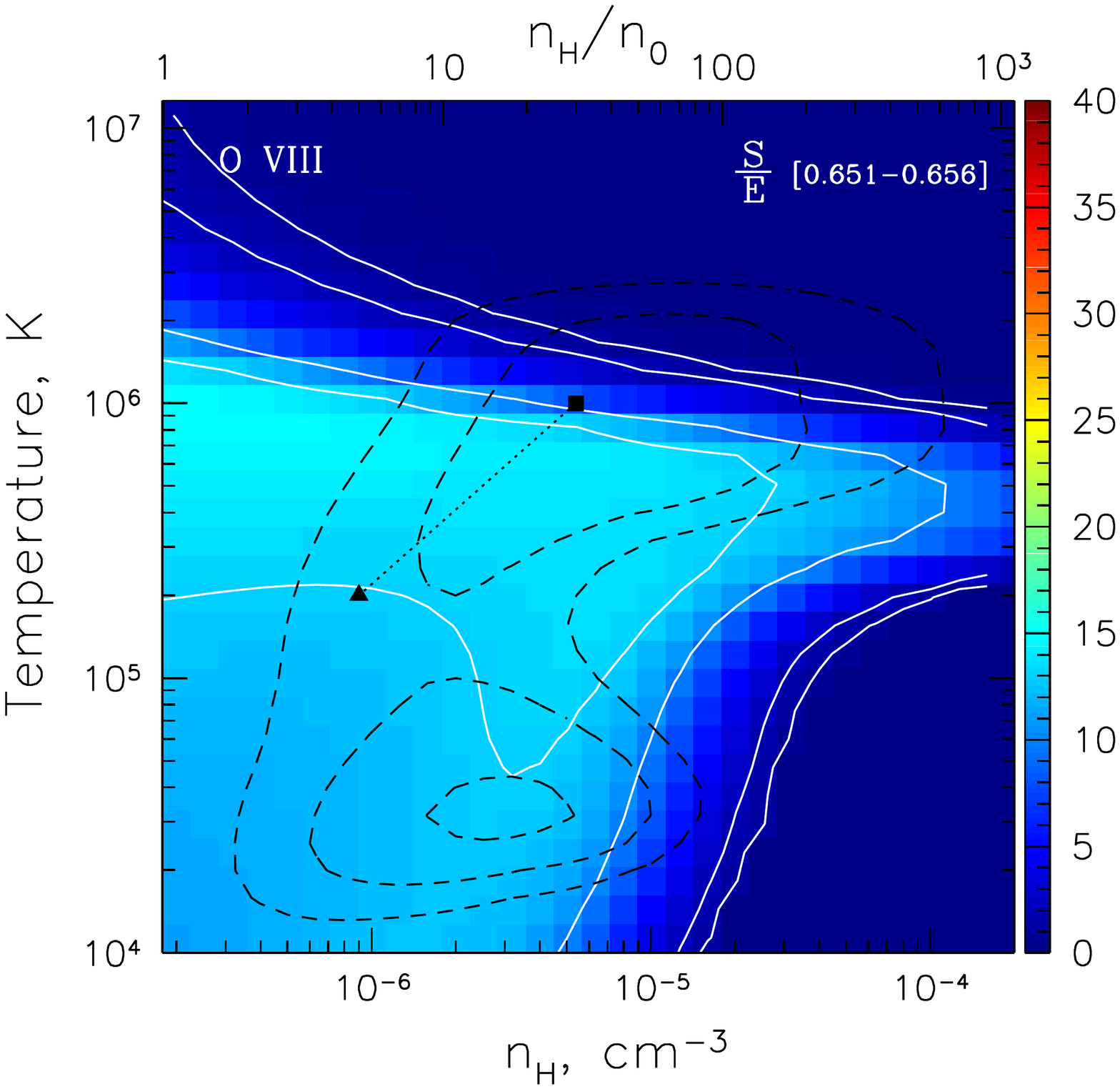}
\includegraphics[bb=50 150 620 700,width=0.31\textwidth]{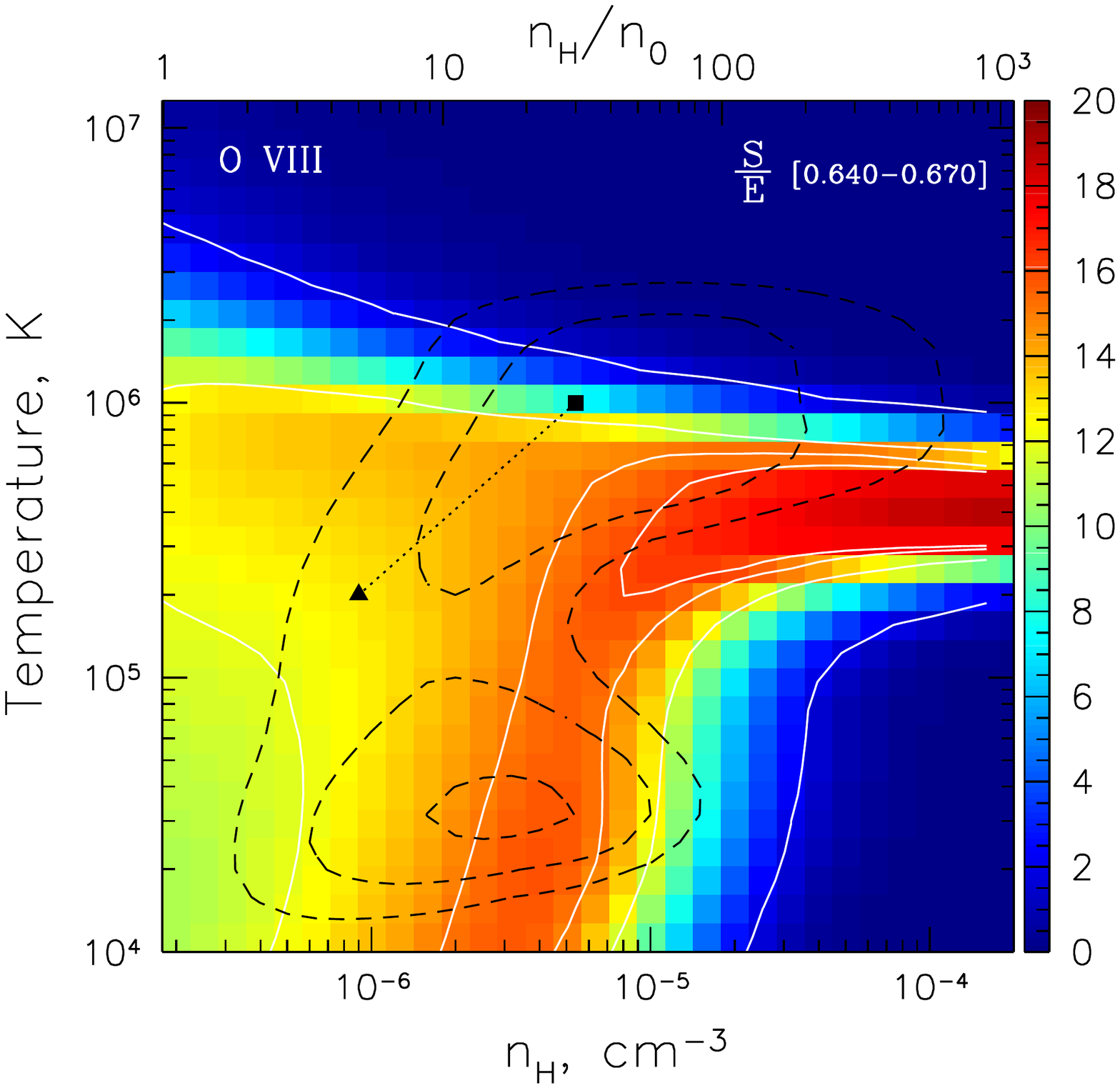}
\includegraphics[bb=50 150 600 700,width=0.31\textwidth]{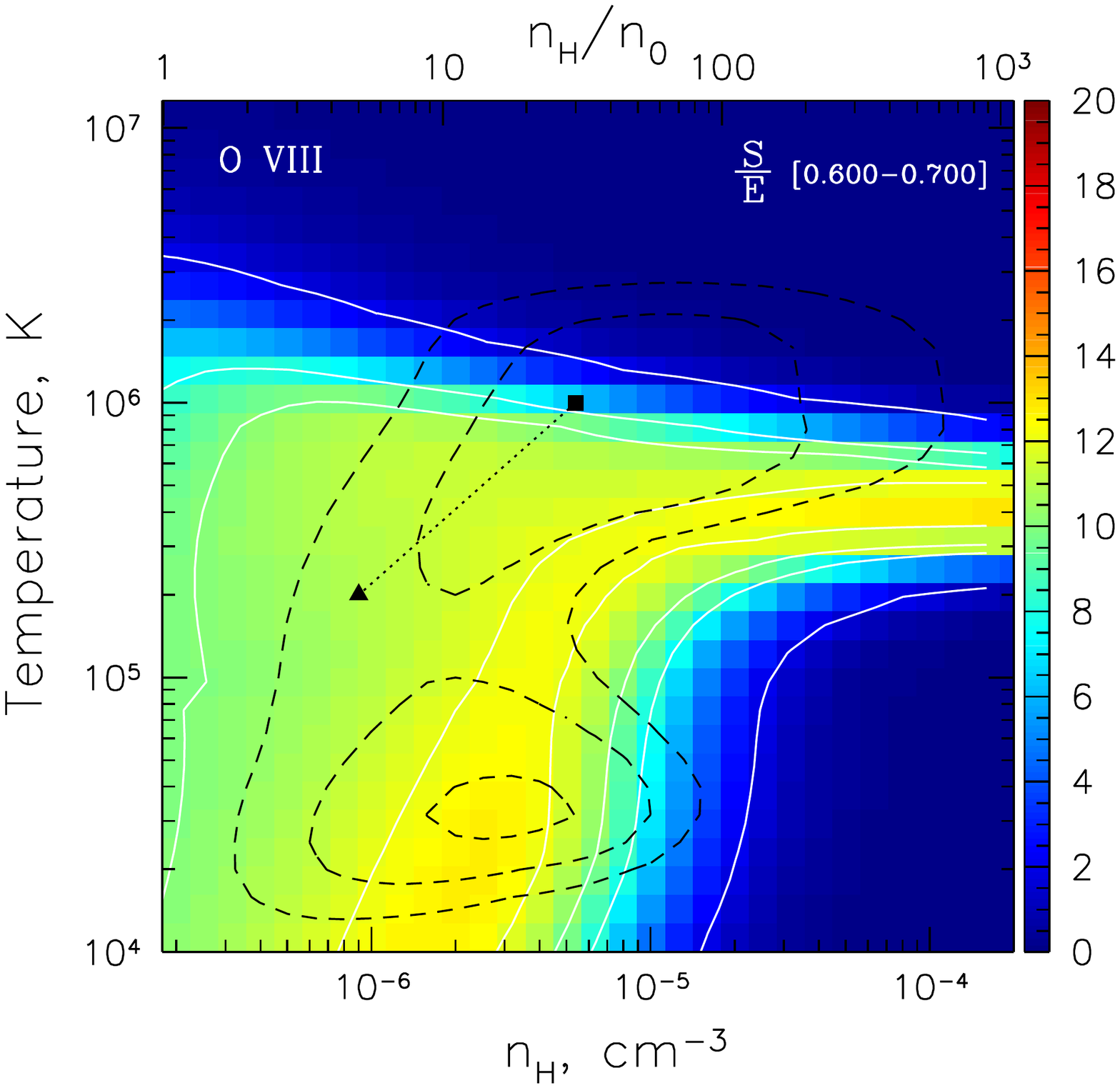}
\caption{{Same as Figure \ref{fig:scemrat}, but for the 5-eV, 30-eV, and 100-eV-wide windows centred on the brightest O VIII line. The spectral bands are  0.651-0.656 keV (left panel, white solid contours at 1, 2, 10 and 13), 0.640-0.670 keV (middle panel, white solid contours at 2, 12, 15 and 16) and 0.600-0.700 keV (right panel, white solid contours at 2, 8, 10 and 12). Note the different colorbar scale of the left panel. The black dashed contours and marks are the same as in Figure \ref{fig:scemrat}.}}
\label{fig:scemrat_oviii}
\end{figure*}

\begin{figure*}
\centering
\includegraphics[bb=50 150 620 700,width=0.31\textwidth]{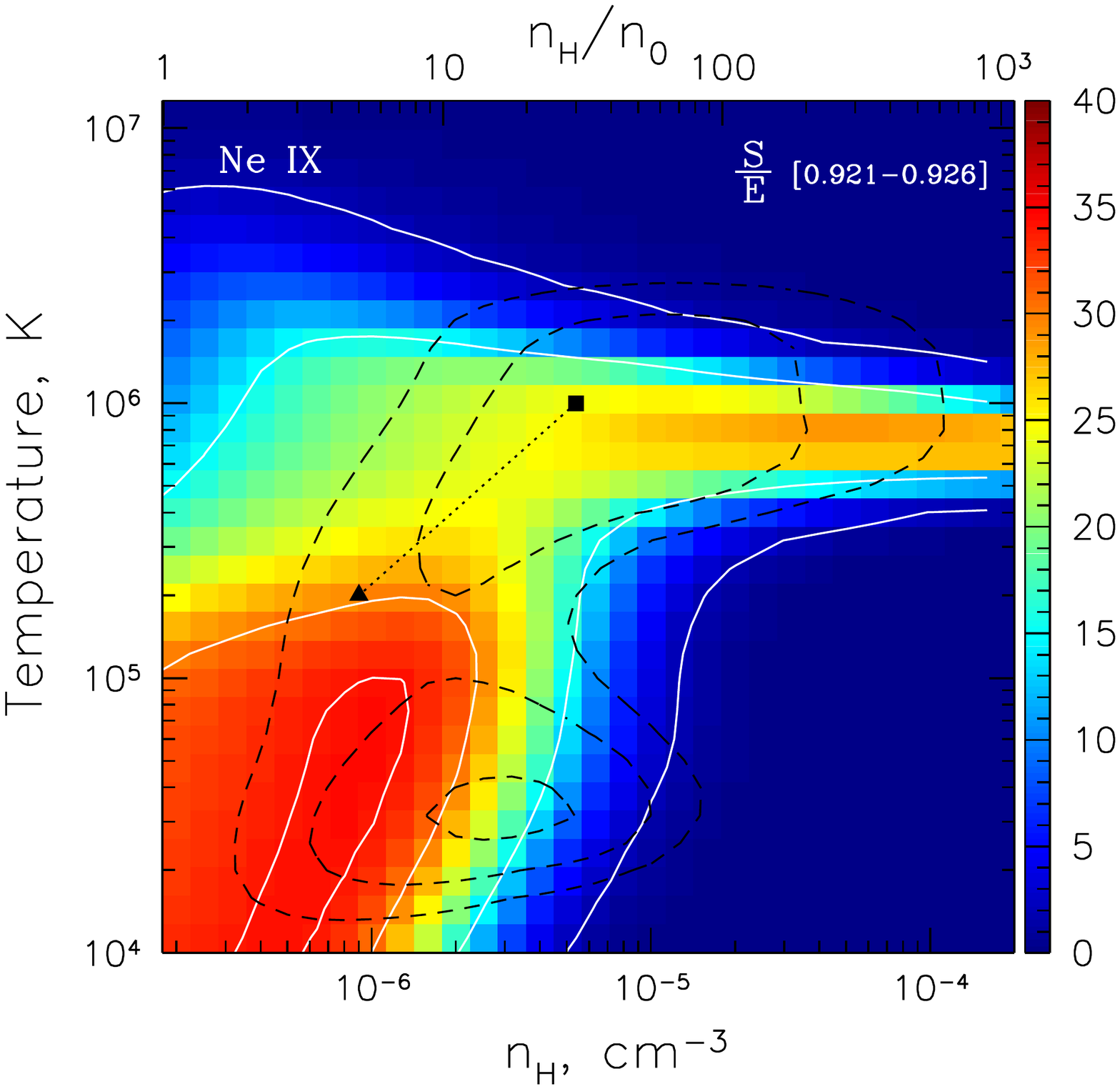}
\includegraphics[bb=50 150 620 700,width=0.31\textwidth]{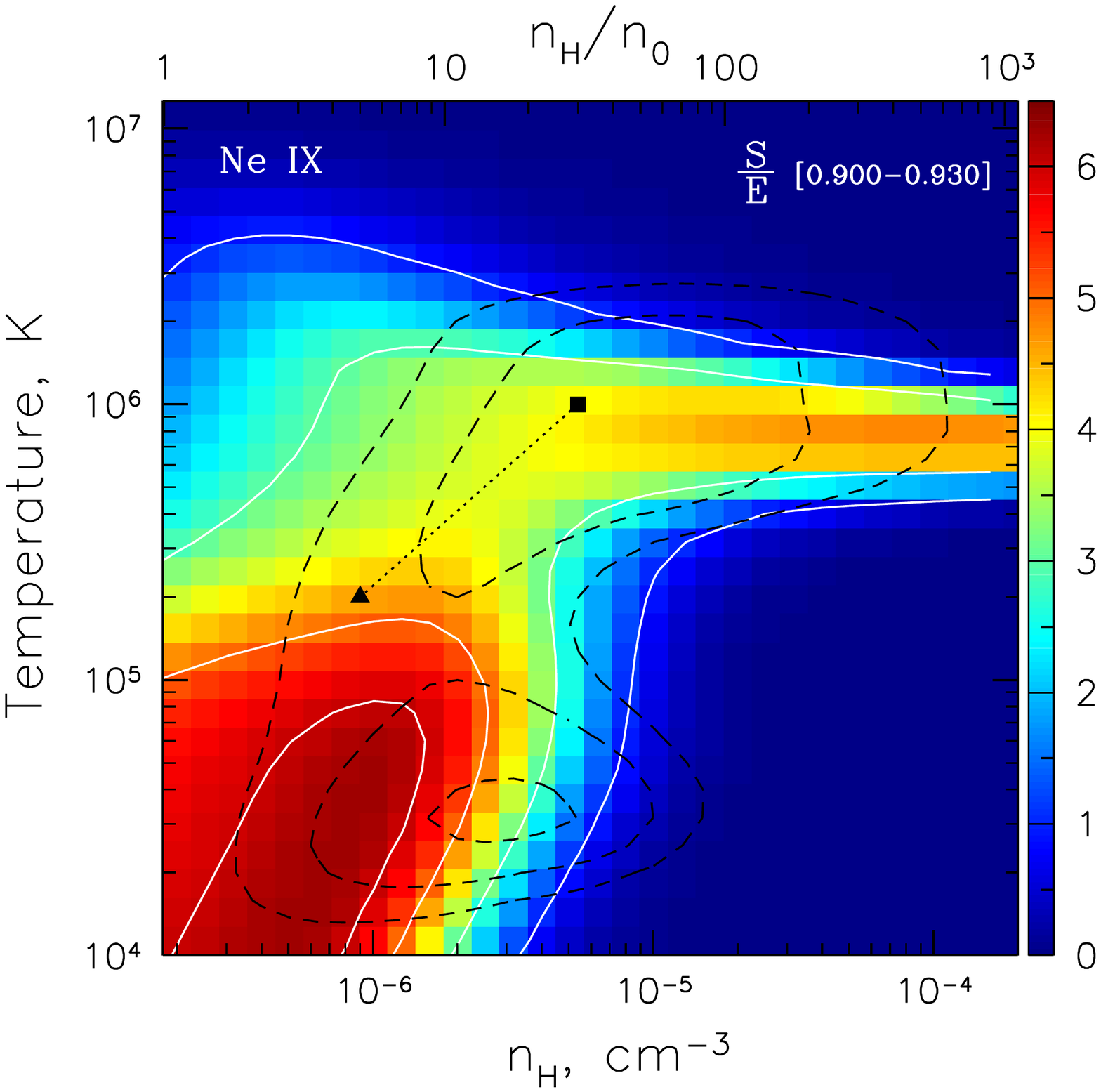}
\includegraphics[bb=50 150 600 700,width=0.31\textwidth]{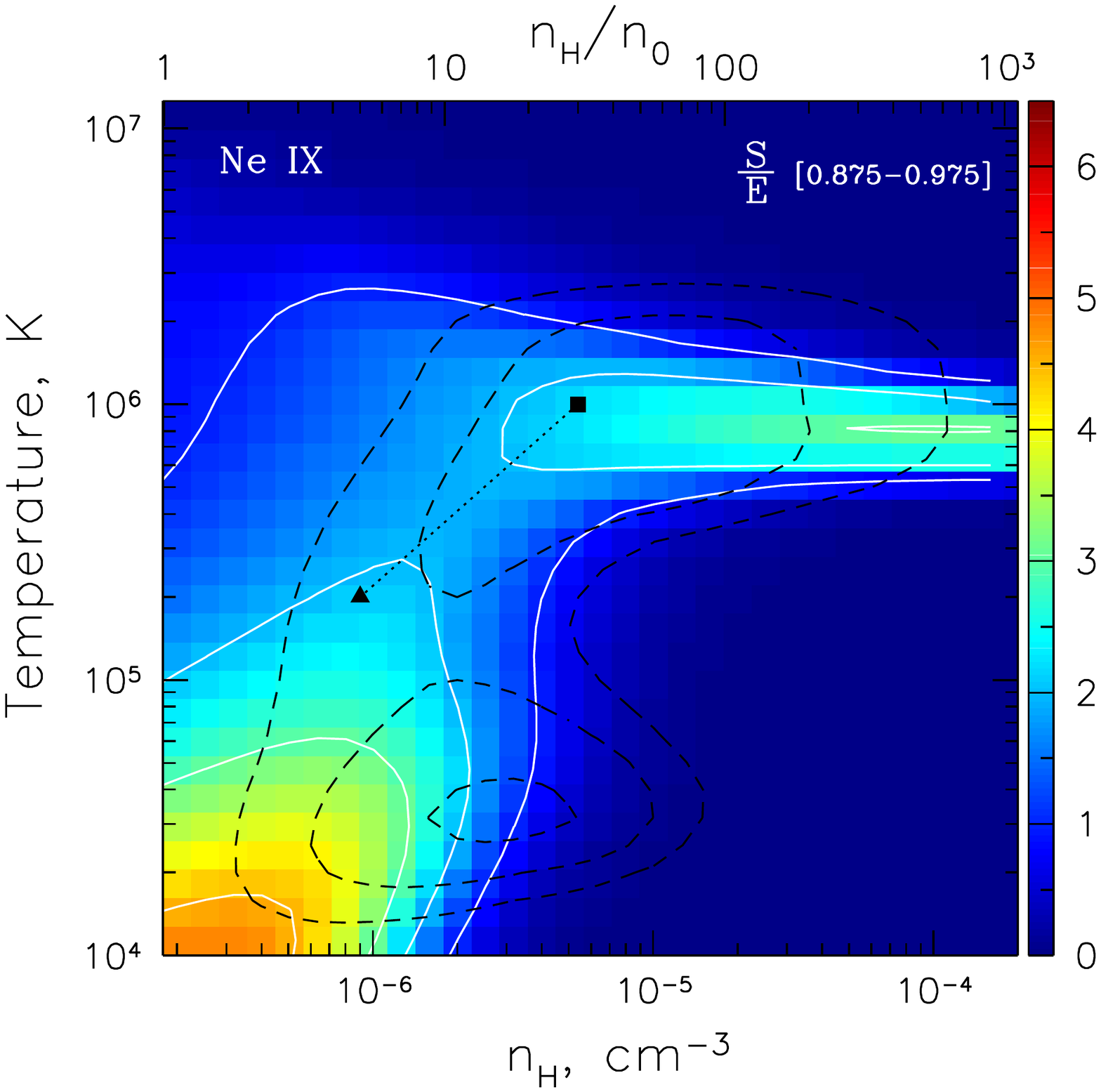}
\caption{{Same as Figure \ref{fig:scemrat}, but for the 5-eV, 30-eV, and 100-eV-wide windows centred on the resonant Ne IX line. The spectral bands are  0.921-0.926 keV (left panel, white solid contours at 2, 15, 30 and 34), 0.900-0.930 keV (middle panel, white solid contours at 1, 3, 5 and 6) and 0.875-0.975 keV (right panel, white solid contours at 1, 2, 3 and 4.5). Note the different colorbar scale of the left panel. The black dashed contours and marks are the same as in Figure \ref{fig:scemrat}.}}
\label{fig:scemrat_neix}
\end{figure*}

\bibliographystyle{mnras}


\label{lastpage}
\end{document}